\documentclass[12pt,twoside]{article}

\usepackage{pslatex}   
\usepackage{amsmath}
\usepackage{amsxtra}
\usepackage{amscd}
\usepackage{amsthm}
\usepackage{amsfonts}
\usepackage{amssymb}
\usepackage{eucal}
\usepackage{epsf}
\usepackage{cite}
\usepackage{a4}

\def\({\left(}
\def\){\right)}

\newcommand{\bea}{\begin{eqnarray}}
\newcommand{\ena}{\end{eqnarray}}
\newcommand{\be}{\begin{eqnarray*}}
\newcommand{\en}{\end{eqnarray*}}
\newcommand{\ba}{\begin{array}}
\newcommand{\ea}{\end{array}}
\renewcommand{\o}[1]{\overline{#1}\,}

\newcommand{\bbA}{\mathbb{A}}
\newcommand{\bbB}{\mathbb{B}}
\newcommand{\bbC}{\mathbb{C}}
\newcommand{\bbD}{\mathbb{D}}
\newcommand{\bQ}{\mathbf{Q}}

\newcommand{\bX}{\mathbf{X}}
\newcommand{\bJ}{\mathbb{J}}
\newcommand{\bR}{\mathbb{R}}

\newenvironment{tenumerate}{
  \begin{enumerate}
  
  }{\end{enumerate}}
\newcommand{\bi}{\begin{tenumerate}}
\newcommand{\ei}{\end{tenumerate}}
\newcommand{\isoto}[1][]%
{{\mathop{\buildrel{\sim}\over\longrightarrow}\limits_{#1}}}

\newcommand{\la}{\lambda}
\newcommand{\g}{\gamma}
\newcommand{\al}{\alpha}
\newcommand{\e}{\epsilon}

\newcommand{\ve}{\varepsilon}
\newcommand{\z}{\zeta}

\newcommand{\om}{\omega}

\newcommand{\bb}{\mathbf{b}}
\newcommand{\bc}{\mathbf{c}}

\newcommand{\bT}{\mathbb{T}}

\def\beq{\begin{equation}}
\def\eeq{\end{equation}}
\def\bea{\begin{eqnarray}}
\def\eea{\end{eqnarray}}

\def\beann{\begin{eqnarray*}}
\def\eeann{\end{eqnarray*}}

\let\a=\alpha \let\be=\beta \let\g=\gamma \let\de=\delta
\let\e=\varepsilon \let\z=\zeta \let\h=\eta 

  \let\la=\lambda \let\m=\mu
 \let\x=\xi \let\p=\pi \let\r=\rho \let\s=\sigma
\let\om=\omega \let\ps=\psi
\let\ph=\varphi   
\let\Om=\Omega  
  \let\D=\Delta

\let\qd=\quad  

\def\epp{\, .}
\def\epc{\, ,}

\theoremstyle{plain}

\newtheorem{proposition}{Proposition}

\newtheorem*{corollary*}{Corollary}

\newtheorem*{conjecture*}{Conjecture}

\theoremstyle{definition}

\def\2{\frac{1}{2}} \def\4{\frac{1}{4}}

\def\6{\partial}

\def\+{\dagger}

\def\<{\langle} \def\>{\rangle}

\def\CH{{\cal H}} 
\def\CO{{\cal O}}

\def\CV{{\cal V}} \def\CW{{\cal W}}

\def\i{{\rm i}}

\def\rd{{\rm d}}
\def\re{{\rm e}}

\DeclareMathOperator{\sh}{sh}
\DeclareMathOperator{\ch}{ch}
\DeclareMathOperator{\tgh}{th}

\DeclareMathOperator{\cth}{cth}

\DeclareMathOperator{\tr}{tr}

\DeclareMathOperator{\sign}{sign}
\DeclareMathOperator{\End}{End}
\DeclareMathOperator{\id}{id}
\DeclareMathOperator{\res}{res}

\DeclareMathOperator{\sing}{sing}

\def\fa{\mathfrak{a}}

\def\faq{\overline{\mathfrak{a}}}

\renewcommand{\appendix}{%
   \renewcommand{\section}{
        \secdef\Appendix\sAppendix}%
   \setcounter{section}{0}%
   \renewcommand{\thesection}{\Alph{section}}%
   \renewcommand{\theequation}{\thesection.\arabic{equation}}%
}

\newcommand{\Appendix}[2][?]{%
     \refstepcounter{section}%
     \setcounter{equation}{0}%
     \addcontentsline{toc}{appendix}%
          {\protect\numberline{\appendixname~\thesection} #1}%
     \vspace{\baselineskip}%
     {\noindent\Large\bfseries\appendixname\ \thesection: #2\par}%
     \sectionmark{#1}\vspace{\baselineskip}}

\newcommand{\sAppendix}[1]{%
     {\noindent\large\bfseries\appendixname\:: #1\par}%
     \sectionmark{#1}\vspace{\baselineskip}}

\renewcommand{\tilde}{\widetilde}
\begin{document} 

\begin{center}

{\Large {\bf Factorization of the finite temperature correlation
functions of the XXZ chain in a magnetic field\\[4ex]}}

{\large Herman E. Boos\footnote{e-mail: boos@physik.uni-wuppertal.de,
on leave of absence from 
Skobeltsyn Institute of Nuclear Physics, 
MSU, 119992, Moscow, Russia
},
Frank G\"{o}hmann\footnote{e-mail: goehmann@physik.uni-wuppertal.de},
Andreas Kl\"umper\footnote{e-mail: kluemper@physik.uni-wuppertal.de}%
\\[1ex]

Fachbereich C -- Physik, Bergische Universit\"at Wuppertal,\\
42097 Wuppertal, Germany\\[2ex]

and\\[2ex] Junji Suzuki\footnote{
e-mail: sjsuzuk@ipc.shizuoka.ac.jp
}\\[1ex]
Department of Physics, Faculty of Science, Shizuoka University,\\
Ohya 836, Suruga, Shizuoka, Japan}

\thispagestyle{empty}

\vspace{20mm}

{\large {\bf Abstract}}

\end{center}

\begin{list}{}{\addtolength{\rightmargin}{9mm}
               \addtolength{\topsep}{-5mm}}
\item
We present a conjecture for the density matrix of a finite segment of
the XXZ chain coupled to a heat bath and to a constant longitudinal
magnetic field. It states that the inhomogeneous density matrix,
conceived as a map which associates with every local operator its
thermal expectation value, can be written as the trace of the
exponential of an operator constructed from weighted traces of the
elements of certain monodromy matrices related to
$U_q (\widehat{\mathfrak{sl}}_2)$ and only two transcendental functions
pertaining to the one-point function and the neighbour correlators,
respectively. Our conjecture implies that all static correlation
functions of the XXZ chain are polynomials in these two functions
and their derivatives with coefficients of purely algebraic origin.
\\[2ex]
\end{list}

\clearpage

\section{Introduction}

The past two decades have seen significant progress in the
understanding of the correlation functions of local operators in
spin-$\frac{1}2$ chains. This report is about the extention of
recent results for the ground state correlators of the XXZ chain,
surveyed below, to finite temperatures.

The development 
was initiated with the derivation
of a multiple integral formula for the density matrix of the
XXZ chain by the Kyoto school \cite{JMMN92,JiMi96,JiMi95} which
relies on the bosonization of $q$-vertex operators and on the
$q$-Knizhnik-Zamolodchikov equation \cite{Smirnov92b,FrRe92}. 
An alternative derivation of the multiple integral formula was found
in \cite{KMT99b}. It is based on the algebraic Bethe ansatz and made
it possible to include a longitudinal magnetic field. 

The multiple integral formulae, however, turned out to be numerically
inefficient.  They were hence not much used before it was
realized \cite{BoKo01} that they may be calculated by hand, at least
in principle. This result  
generalized after many years
Takahashi's curious formula \cite{Takahashi77} for the next-to-nearest
neighbour correlator and inspired a series of works devoted to
the explicit calculation of short-distance correlators in the XXX
\cite{BoKo02,BKNS02,BST05,SSNT03,SaSh05,SST05} and XXZ chains
\cite{KSTS03,KSTS04,TKS04}. It further triggered a deep investigation
into the mathematical structure of the inhomogeneous density matrix
of the XXZ chain, which was started in \cite{BKS03,BKS04a,BKS04c} and
still continues \cite{BJMST04a,BJMST04b,BJMST05a,BJMST05b,BJMST06,%
BJMST06b}.

In \cite{BJMST04a} a minimal set of equations that determines the
inhomogeneous density matrix was derived and was termed the reduced
$q$-Knizhnik-Zamolodchikov (rqKZ) equation. The rqKZ equation made it
possible to prove that the correlation functions of the inhomogeneous
XXX model depend on a single transcendental function which is basically
the two-spinon scattering phase. This was generalized to the XXZ and
XYZ models in \cite{BJMST04b,BJMST05a}, where further transcendental
functions were needed.

A new `exponential form' of the density matrix was derived in
\cite{BJMST05b} and \cite{BJMST06} for which the homogeneous (physical)
limit can be taken directly. 
The most recent papers \cite{BJMST06b,BJMST07app} aimed at understanding
how the exponential formula works in the `free fermion' XX limit. This
led to a novel formulation also for generic~$q$. A crucial tool was a
disorder field acting on half of the infinite lattice with `strength'
$\al$. It regularized the problem further and simplified the
exponential formula in a way that the exponent depends only on a
single transcendental function $\om$ and on special operators $\bb$
and $\bc$ resembling annihilation operators of (Dirac) fermions.

From the above studies we observe the following. In the inhomogeneous
case the multiple integrals reduce to polynomials in a small number
of different single integrals related to the correlation functions
of only nearest-neighbouring lattice sites. These constitute a set
of transcendental functions which determine what we call the `physical
part' of the problem. The coefficients of the polynomials are rational
functions of the inhomogeneity parameters. They are constructed from
various $L$-operators related to the symmetry of the models and
constitute the `algebraic part'. We call such type of separation
of the problem into a finite physical part and into an algebraic part 
`factorization', since it can be traced back to the factorization
of multiple integrals into single integrals. We believe that
factorization is a general feature of integrable models (for a similar
phenomenon  in the form factors for the Ising model see \cite{BHMMOZ06}). 

A generalization of the integral formula for the density matrix 
of the XXZ chain to finite temperature and magnetic field was derived
in \cite{GKS05b,GKS05,GHS05} by combining the techniques developed in
\cite{KMT99b} with the finite temperature formalism of
\cite{Suzuki85,SuIn87,SAW90,Kluemper92,Kluemper93}. 
Remarkably, the form of the multiple
integrals for the density matrix elements is the same in all known
cases. The physical parameters (temperature $T$, magnetic field $h$, chain
length $L$) enter only indirectly through an auxiliary function which is
defined as a solution of a non-linear integral equation.

The auxiliary function enters into the multiple integrals as a weight
function. This implies that the factorization technique developed for
the ground state correlators in \cite{BoKo01} does not work any longer.
In our previous work \cite{BGKS06} we nevertheless obtained a
factorization of the correlation functions of up to three neighbouring
sites in the XXX model at arbitrary $ T, h $
by implicit use of a certain
integral equation. Comparing the factorized forms with
the known results for the ground state we could conjecture an exponential
formula for the special case of $T>0$  but  $h=0$.
 Surprisingly, the formula shares the same algebraic
part with its $T=0$ counterpart; one only has to replace the
transcendental function by its finite temperature generalization.
The results 
 easily  translated into similar results for
the ground state of the system of finite length \cite{DGHK07}.

In this work we extend our analysis 
to the periodic XXZ chain
\begin{equation} \label{xxzham}
     \CH_N = J \sum_{j=-N+1}^N \Bigl( \s_{j-1}^x \s_j^x
             + \s_{j-1}^y \s_j^y + \D (\s_{j-1}^z \s_j^z - 1) \Bigr)
\end{equation}
in the antiferromagnetic regime ($J > 0$ and $\D = \ch (\h) > - 1$)
and in the thermodynamic limit ($L=2N \rightarrow \infty$). We identify
an appropriate set of basic functions describing the neighbour
correlators in the inhomogeneous case. The algebraic part of the
problem without magnetic field is neatly formulated in terms of
the operators $\bb$ and $\bc$ as in the ground state case. The meaning
of the disorder parameter $\a$, necessary for the construction of
these operators, is yet to be understood for finite temperatures.
It, however, naturally modifies one of our auxiliary functions, the
density function $G$ and allows us to reduce the number of basic
functions characterizing the physical part from two to one.

Still, we go one important step further. We extend our conjectured 
exponential formula for the (finite temperature) density matrix
such as to include the magnetic field. At first sight, this may
seem to require only trivial modifications, as the Hamiltonian commutes
with the Zeeman term.
The magnetic field, however,
breaks the $U_q(\widehat{\mathfrak{sl}}_2)$ symmetry and, as far as the
factorization of the integrals is concerned, brings about serious
difficulties even for the ground state correlator problem.
For this reason an essential modification of the operator in the
exponent of our exponential formula is required which leads to
novel formulae even in the zero temperature limit. 
The prescription is, however,  remarkably simple.
We have to add
a term whose algebraic part is determined by a new operator $\mathbf{H}$,
such that the operator in the exponent is now a sum of two
ingredients. One is formally identical to the operator already
present at vanishing magnetic field, the other one is constructed from
$\mathbf{H}$ (note that  even the former part is not independent of 
the field; it includes transcendental functions which are even
functions of $h$).

We finally point out a simplification compared to the ground state
case, particularly relevant at finite magnetic field. Although  
we are dealing with highly nontrivial functions, all correlation
functions  should simplify in the vicinity of $T = \infty$.
Thus, the high temperature expansion technique can be applied to 
the multiple integral formulae at $T>0$ as was shown
in \cite{TsSh05,Tsuboi07}. We use this in order to test
our conjecture for the exponential form of the density matrix.

Our paper is organized as follows. In section \ref{sec:multint} we
recall the definition of the density matrix and the multiple integral
formulae. In section \ref{sec:basefun} we describe the basic functions
that determine the physical part of the correlation functions. Our
main result is presented in section \ref{sec:corfactor} (see eqs.\
(\ref{main2})-(\ref{Omega1and2})). It is a conjectured exponential
formula for the density matrix of the XXZ chain at finite temperature
and magnetic field. Section \ref{sec:examples} is devoted to the
simplest examples of correlation functions, the cases of $n = 1, 2, 3$,
for which we show novel explicit formulae. In section \ref{sec:concl}
we summarize and discuss our results. Appendix \ref{app:proofs} contains
the proofs of two formulae needed in the main body of the paper,
appendix \ref{app:doubleint} a derivation of the factorized form of
the density matrix for $n = 2$ directly from the double integrals, and
appendix \ref{app:hte} a short description of the high-temperature
expansion technique.

\section{Multiple integral representation of the density matrix}
\label{sec:multint}
Let us recall the definition of the density matrix of a chain
segment of length $n$. We would like to take into account a
longitudinal magnetic field $h$ which couples to the conserved
$z$-component
\begin{equation} \label{totalsz}
     S_N^z = \2 \sum_{j=-N+1}^N \s_j^z
\end{equation}
of the total spin. Then the statistical operator of the equilibrium
system at temperature $T$ is given by
\begin{equation} \label{statop}
     \r_N (T,h) = \frac{\re^{- \frac{\CH_N}{T} + \frac{h S_N^z}{T}}}
                       {\tr_{-N+1, \dots, N} \,
		        \re^{- \frac{\CH_N}{T} + \frac{h S_N^z}{T}}}
			\epp
\end{equation}
From this operator we obtain the density matrix of a chain segment
of length $n$ by tracing out the complementary degrees of freedom,
\begin{equation} \label{densmatgen}
     D_n (T,h|N) = \tr_{-N+1, \dots, -1, 0, n+1, n+2, \dots, N} \,
                   \r_N (T,h) \epc \qd n = 1, \dots, N \epp
\end{equation}
The density matrix $D_n (T,h|N)$ encodes the complete equilibrium
information about the segment consisting of sites $1, \dots, n$ which
means that every operator $\CO$ acting non-trivially at most on sites
$1, \dots, n$ has thermal expectation value
\begin{equation} \label{oexp}
     \< \CO \>_{T, h}  = \tr_{1, \dots, n} \bigl( D_n (T,h|N) \CO \bigr)
                         \epp
\end{equation}

\begin{figure}

\begin{center}

\epsfxsize 14cm
\epsffile{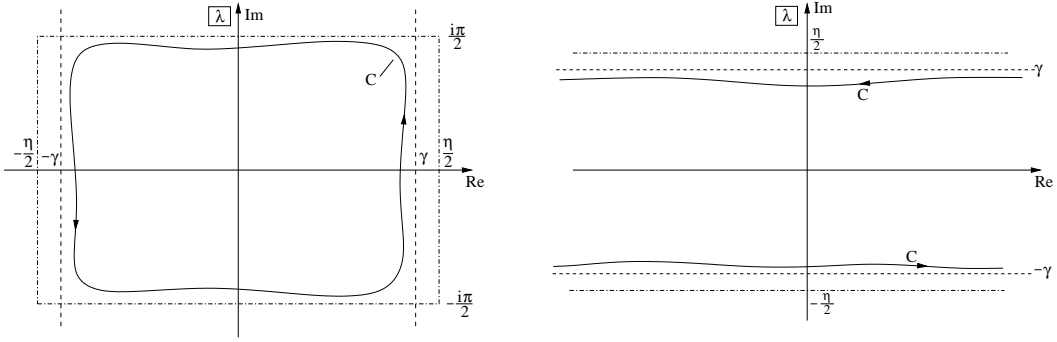}

\caption{\label{fig:cancon} The canonical contour ${\cal C}$ for the
off-critical regime $\D > 1$ (left) and for the critical regime
$- 1 < \D < 1$ (right).}
\end{center}

\end{figure}

We know a multiple integral representation for the density matrix
(\ref{densmatgen}) in two limiting cases, the thermodynamic limit
$N \rightarrow \infty$ \cite{GKS05b,GHS05} and the zero temperature
and zero magnetic field limit \cite{DGHK07}. For the two limits
we shall employ the notation
\begin{equation} \label{densthn}
     D_n (T,h) = \lim_{N \rightarrow \infty} D_n (T,h|N) \epc \qd
     D_n (N) = \lim_{T \rightarrow 0} \lim_{h \rightarrow 0} D_n (T,h|N)
               \epp
\end{equation}
These two density matrices are conveniently described in terms of
the canonical basis of endomorphisms on ${(\mathbb C}^2)^{\otimes n}$
locally given by $2 \times 2$ matrices $e^\a_\be$, $\a, \be = \pm$,
with a single non-zero entry at the intersection of row $\be$ and
column $\a$,
\begin{equation} \label{densbase}
     D_n (T,h) = {D_n\,}_{\a_1, \dots, \a_n}^{\be_1, \dots, \be_n} (T, h)
                 \, {e_1}_{\a_1}^{\be_1} \dots {e_n}_{\a_n}^{\be_n}
		 \epc \qd
     D_n (N) = {D_n\,}_{\a_1, \dots, \a_n}^{\be_1, \dots, \be_n} (N)
               \, {e_1}_{\a_1}^{\be_1} \dots {e_n}_{\a_n}^{\be_n} \epc
\end{equation}
where we assume implicit summation over all $\a_j, \be_k = \pm$.
We further regularize the density matrices by introducing a set of
parameters $\la_1, \dots, \la_n; \a$ in such a way that
\begin{subequations}
\begin{align} \label{densinhTh}
     D_n (T,h) & = \lim_{\la_1, \dots, \la_n \rightarrow 0} \:
                   \lim_{\a \rightarrow 0}
                   {D_n\,}_{\a_1, \dots, \a_n}^{\be_1, \dots, \be_n}
		          (\la_1, \dots, \la_n |T, h; \a) \,
		          {e_1}_{\a_1}^{\be_1} \dots {e_n}_{\a_n}^{\be_n}
                          \epc \\ \label{densinhN}
     D_n (N) & = \lim_{\la_1, \dots, \la_n \rightarrow \h/2} \:
                 \lim_{\a \rightarrow 0}
                 {D_n\,}_{\a_1, \dots, \a_n}^{\be_1, \dots, \be_n}
	                (\la_1, \dots, \la_n |N; \a) \,
		        {e_1}_{\a_1}^{\be_1} \dots {e_n}_{\a_n}^{\be_n}
			\epp
\end{align}
\end{subequations}
From here on we shall concentrate on the temperature case
(\ref{densinhTh}). Later we will indicate the modifications necessary
for (\ref{densinhN}). We call ${D_n\,}_{\a_1, \dots, \a_n}^%
{\be_1, \dots, \be_n} (\la_1, \dots, \la_n |T, h; \a)$ the inhomogeneous
density matrix element with inhomogeneity parameters $\la_j$. For
$\a = 0$ it has a clear interpretation in terms of the six-vertex
model with spectral parameters $\la_1, \dots, \la_n$ on $n$ consecutive
vertical lines \cite{GKS05}. For $h, T = 0$ the variable $\a$ can be
interpreted as a disorder parameter \cite{JiMi95}. In the general
case we simply define the inhomogeneous density matrix element by
the following multiple integral,
\begin{align} \label{densint}
     {D_n\,}_{\a_1, \dots, \a_n}^{\be_1, \dots, \be_n}
           (\la_1, \dots, & \la_n |T, h; \a)
        = \notag \\ \de_{s,m-s'}
	  \biggl[ \prod_{j=1}^s
	     \int_{\cal C} & \frac{\rd \om_j \: \re^{- \a \h}}
	                          {2 \p \i (1 + \fa (\om_j))}
	     \prod_{k=1}^{x_j - 1} \sh(\om_j - \la_k - \h)
	     \prod_{k = x_j + 1}^n \sh(\om_j - \la_k) \biggr] \notag \\
          \biggl[ \prod_{j = s + 1}^n
	     \int_{\cal C} & \frac{\rd \om_j \: \re^{\a \h}}
	                          {2 \p \i (1 + \faq (\om_j))}
	     \prod_{k=1}^{x_j - 1} \sh(\om_j - \la_k + \h)
	     \prod_{k = x_j + 1}^n \sh(\om_j - \la_k) \biggr] \notag \\ &
        \frac{\det[ - G(\om_j, \la_k; \a)]}
	     {\prod_{1 \le j < k \le n}
	         \sh(\la_k - \la_j) \sh( \om_j - \om_k - \h)} \epp
\end{align}
Here $s$ is the number of plus signs in the sequence $(\a_j)_{j=1}^n$,
and $s'$ is the number of minus signs in the sequence $(\be_j)_{j=1}^n$.
The factor $\de_{s,m-s'}$ reflects the conservation of the $z$-component
of the total spin. For $j = 1, \dots, s$ the variable $x_j$ denotes the
position of the $j$th plus sign in $(\a_j)_{j=1}^n$ counted from the
right. For $j = s + 1, \dots, n$ it denotes the position of $(j - s)$th
minus sign in $(\be_j)_{j=1}^n$. The integration contour depends on
$\h$. We show it in figure \ref{fig:cancon}. This contour will also
appear in the integral equations which determine the transcendental
functions $\fa$, $\faq$ and $G$ and in the definition of the
special functions in the next section that determine the physical
part in the factorized form of the correlation functions. For this
reason we call it the canonical contour.

The integral equation for $\fa$ is non-linear,
\begin{equation} \label{nlie}
     \ln \fa (\la) = - \frac{h}{T}
                     - \frac{2J \sh^2 (\h)}{T \sh(\la) \sh(\la + \h)}
                     - \int_{\cal C} \frac{\rd \om}{2 \p \i} \,
                       \frac{\sh (2 \h) \ln (1 + \fa (\om))}
                            {\sh(\la - \om + \h)
                                    \sh(\la - \om - \h)} \epp
\end{equation}
There is a similar integral equation for $\faq$ (see \cite{GKS04a}),
however, since $\faq = 1/\fa$ we do not need to consider
it here. $\fa$ is usually called the auxiliary function.
The combination $1/(1 + \fa)$ has a natural interpretation as a
generalization of the fermi function to the interacting case
\cite{GoSe05}. Note that the right hand side of equation (\ref{nlie})
is the only place where the thermodynamic variables $T$ and $h$ enter
explicitly into our formulae for correlation functions. They neither
enter explicitly into the multiple integral formula (\ref{densint})
nor into the linear integral equation for $G$ which is
\begin{equation} \label{G}
     G(\la,\mu;\al) =  -\coth(\la-\mu)
                      + \re^{\al\eta} \coth(\la-\mu-\eta)+
                        \int_{\cal C}
			\frac{\rd \om \: G(\om,\mu;\al)}
			     {2\pi i (1+\fa(\om))} K(\la-\om;\al) \epp
\end{equation}
$G$ can be interpreted as a generalized magnetization density (see
\cite{GKS04a}). Compared to our previous definition \cite{GKS04a} we
introduced the additional parameter $\a$ here which also enters the
kernel,
\begin{equation} \label{kernel}
     K(\la;\al) = \re^{\al\eta} \coth(\la-\eta)
                  - \re^{-\al\eta} \coth(\la+\eta) \epp
\end{equation}
An equivalent integral equation for $G$ which uses $\faq$ instead
of $\fa$ and which is sometimes useful is
\begin{equation} \label{G1}
     G(\la,\mu;\al) =  -\coth(\la-\mu)
                      + \re^{-\al\eta} \coth(\la-\mu+\eta) -
                      \int_{\cal C}
		      \frac{\rd \om \: G(\om,\mu;\al)}
		           {2\pi i (1+\o{\fa}(\om))} K(\la-\om;\al) \epp
\end{equation}
Setting $\a = 0$ the function $G(\la,\m;\a)$ turns into the function
$G(\la,\m)$ which played a crucial role in our previous studies
\cite{GKS04a,GKS05,BGKS06}. We have introduced $\a$ in such a way
into (\ref{densint}) and (\ref{G}) that for $T, h = 0$ the multiple
integral representation (\ref{densint}) turns into the finite-$\a$
expression that can be obtained within the $q$-vertex operator
approach of \cite{JiMi95}. Our main motivation for introducing $\a$
into our functions was to enforce compatibility with the formalism
developed in \cite{BJMST06b}, where $\a$ is an important regularization
parameter. The usefulness of this modification will become clear in
section \ref{sec:corfactor}. The parameter $\a$ will allow us to write
our formula for the density matrix in factorized form in a very compact
way.

Let us briefly indicate the changes that are necessary in the finite
length case (\ref{densinhN}). It turns out \cite{DGHK07} that
${D_n\,}_{\a_1, \dots, \a_n}^{\be_1, \dots, \be_n}
(\la_1, \dots, \la_n |N; \a)$ has a multiple integral representation
of the same form as (\ref{densint}), that even the integral
equation for $G$ remains the same and that the only necessary
modification is in the driving term of the non-linear integral
equation (\ref{nlie}), where the physical parameters enter, which
in this case are the length $L = 2N$ of the chain and an arbitrary
twist $\Phi \in [0,2\p)$ of the periodic boundary conditions (for
details see \cite{DGHK07}). The non-linear integral equation for
the finite length case is
\begin{equation} \label{nliefini}
     \ln \fa (\la) = - 2\i \Phi + L \h + L \ln \biggl(
                       \frac{\sh(\la - \frac{\h}{2})}
		            {\sh(\la + \frac{\h}{2})} \biggr)
                     - \int_{\cal C} \frac{\rd \om}{2 \p \i}
                       \frac{\sh (2 \h) \ln (1 + \fa (\om))}
                            {\sh(\la - \om + \h)
                                    \sh(\la - \om - \h)} \epp
\end{equation}
When we derived the multiple integral representation (\ref{densint})
in \cite{GHS05} and \cite{DGHK07} we assumed that the inhomogeneity
parameters $\la_j$ are located inside the integration contour $\cal C$.
This has to be taken into account when calculating the homogeneous
limit in (\ref{densinhN}), where the canonical contour should be first
shifted to $\pm \h/2$.

\section{The basic functions}
\label{sec:basefun}
In this section we describe the functions constituting the `physical
part' of the factorized correlation functions of the XXZ chain at
finite $T$ and $h$. A description of the algebraic part will be given
in the next section. According to our experience the physical part of
the correlation functions can be characterized completely by two
transcendental functions $\ph$ and $\om$.

Let us start with the more simple function
\begin{equation} \label{phi}
      \varphi(\mu;\al) = 1 + \int_{\cal C}
                         \frac{\rd \om \: G(\om,\mu;\al)}
			      {\pi i (1+\fa(\om))} \epp
\end{equation}
This function is related to the magnetization $m(T,h)$ through
$\ph (0;0) = - 2 m(T,h)$ which we expect to belong to the physical
part if the magnetic field is non-zero.

In order to introduce the function $\om$ we first of all define
\begin{equation} \label{psi}
     \psi(\mu_1,\mu_2;\al) = \int_{\cal C}
                             \frac{\rd \om \: G(\om,\mu_1;\al)}
			          {\pi i (1+\fa(\om))}
                             \bigl(- \coth(\om-\mu_2) +
			           \re^{-\al\eta} \coth(\om-\mu_2-\eta)
				   \bigr) \epp
\end{equation}
Those readers who are familiar with our previous work \cite{BGKS06}
will recognize this as the anisotropic and `$\a$-deformed' version
of the function $\psi (\m_1, \m_2)$ introduced there. The function
$\om$ is a modification of $\psi$ obtained by adding and multiplying
some explicit functions,
\begin{equation} \label{om}
     \om(\mu_1,\mu_2;\al) = - \re^{\al(\mu_1-\mu_2)}
                              \psi(\mu_1,\mu_2;\al)
			    - \frac{\re^{\al(\mu_1-\mu_2)}}
			           {2\cosh^2(\frac{\al\eta}2)}
		              K(\mu_1-\mu_2;-\al) \epp
\end{equation}
Here $K(\la;\a)$ is the kernel defined in (\ref{kernel}). The relation
between $\om(\m_1,\m_2;\a)$ and $\psi(\m_1,\m_2;\a)$ is similar to
the relation between $\g(\m_1,\m_2)$ and $\psi(\m_1,\m_2)$
in the isotropic case \cite{BGKS06}. The function $\om$ is closely
related to the neighbour correlators (see appendix \ref{app:doubleint}).
In the critical regime for $T,h\rightarrow 0$ it becomes the function
$\om(\z,\al)$ of the paper \cite{BJMST06b} if we set $\z=e^{\mu_1-\mu_2}$.

An important property which follows from the definitions (\ref{kernel})
and (\ref{psi}) is that
\begin{equation} \label{omexchange}
     \om(\mu_2,\mu_1;-\al) = \om(\mu_1,\mu_2;\al) \epp
\end{equation}
It implies
\begin{equation} \label{omexchange1}
     \om(\mu_2,\mu_1;0) = \om(\mu_1,\mu_2;0)\epc \quad
     \om'(\mu_2,\mu_1;0) = - \om'(\mu_1,\mu_2;0) \epc
\end{equation}
where for later convenience we introduced the somewhat unusual notation
\begin{equation} \label{defomprime}
     \om'(\mu_1,\mu_2;\al) = \6_\a \bigl( \re^{\al(\mu_2-\mu_1)}
                             \om(\mu_1,\mu_2;\al) \bigr) \epp
\end{equation}

At this point we would like to stress that the physical parameters
$T$, $h$ or $N$, respectively, do not enter the definitions of
$\ph$ and $\om$ explicitly. The basic functions defined in this section
are therefore suitable for both, the finite temperature and the
finite length case, the only distinction being the use of different
auxiliary function (\ref{nlie}) and (\ref{nliefini}), respectively.

In the high-temperature limit (see appendix \ref{app:hte}) we observe
that
\begin{equation} \label{ombeta}
     \om(\mu_1,\mu_2;\al) = \frac{\re^{\al (\mu_1-\mu_2)}}2
                            \tanh^2 \left( \frac{\al \eta}2 \right)
			    K(\mu_1-\mu_2;-\al)
			    + \CO \left( \frac 1 T \right) \epp
\end{equation}
Using eq.\ (\ref{ombeta}) we conclude that both functions
$\om(\mu_1,\mu_2;0)$ and $\om'(\mu_1,\mu_2;0)$ do not have zeroth order
terms in their high-temperature expansions
\begin{equation} \label{omom'beta}
     \om(\mu_1,\mu_2;0) = \CO (1/T) \epc \qd
     \om'( \mu_1, \mu_2;0) = \CO (1/T) \epp
\end{equation}
The same is true for the function $\varphi$,
\begin{equation} \label{phibeta}
     \varphi(\mu;\al) = \CO (1/T) \epp
\end{equation}

We mention the properties of these functions for $\alpha=0$ with respect
to reversal of the magnetic field; $\varphi(\mu;0)$ is an odd function
of $h$, $\psi(\mu_1,\mu_2;0)$ and $\partial_{\al}
\psi(\mu_1,\mu_2;\al)|_{\al=0}$ are even. These properties will be
implicitly used below. The proof relies on the simple fact that the
quantum transfer matrix (or its slight generalization, see below)
associated to the present model respects the spin reversal symmetry,
and therefore the eigenvalues are even functions of $h$.

Once this is realized, the proof for $\varphi(\mu;0)$ is rather obvious. 
One only has to remember the relation between $\varphi(\mu;0)$ and the
largest eigenvalue $\Lambda(\mu)$ of the quantum transfer matrix,
\begin{equation}
\varphi(\mu;0) = T \frac{\partial}{\partial h} \ln \Lambda(\mu) \epp
\end{equation}
The above argument then implies that $\varphi(\mu;0)$ is odd with
respect to $h$.

The proof for $\psi(\mu_1,\mu_2;0)$ is less obvious. We first of all
introduce a generalized system. Consider an `alternating' inhomogeneous
transfer matrix. In the framework of the quantum transfer matrix, we
associate spectral parameters in alternating manner $(u, -u, u,-u
\cdots)$ to 2${\cal N}$ vertical bonds, while keeping the spectral
parameter on the horizontal axis fixed as $\mu_2$. Next we add
$2{\cal M }$ vertical bonds and associate with them spectral parameters
again in alternating manner, $(u'+\mu_1, \mu_1-u', u'+\mu_1, \mu_1-u',
\cdots)$. We then take the limit ${\cal N}, {\cal M} \rightarrow
\infty$ under the fine tuning, $ 2 u{\cal N} = 2\beta J \sh \eta,
2 u'{\cal M} = -2 \delta J \sh \eta$. Note that the original system
is recovered by taking $\delta=0$. By neglecting the term depending
on the overall normalization, one obtains the following expression for
the modified largest eigenvalue $\Lambda(\mu_2, \m_1)$ of the generalized
quantum transfer matrix,
\begin{equation} \label{generalizedLam}
     \ln \Lambda(\mu_2, \m_1) = -\frac{\beta h}{2} -
        \int_{\cal C} \frac{d\omega}{2\pi i}
	e(\om-\mu_2) \ln (1+\bar{\fa}(\om, \m_1)) \epc \qquad
     e(\la):= \frac{\sh (\h)}{\sh(\la) \sh(\la-\h)} \epp
\end{equation}
The modified auxiliary functions $\fa(\om, \m_1), \bar{\fa}(\om, \m_1)$
satisfy equations similar to (\ref{nlie}), and the equation for the
latter is relevant here,
\begin{equation} \label{nliedelta}
     \ln \bar{\fa} (\la, \m_1) =  \frac{h}{T}
                     - \frac{2J \sh(\h)}{T} e(\la)
		     + 2\delta J \sh(\h)  e(\la-\m_1) 
                     +\int_{\cal C} \frac{\rd \om}{2 \p \i} \,
                       K(\la-\om;0) \ln (1 + \bar{\fa }(\om, \m_1)) \epp
\end{equation}
We take the derivative of both sides of (\ref{nliedelta}) with respect
to $\delta$,
\begin{equation} \label{sigmalinear}
   \sigma(\la,\m_1) =
                    2 J \sh(\h)  e(\la-\m_1) 
                     +\int_{\cal C} \frac{\rd \om}{2 \p \i} \,
                       K(\la-\om;0)  \frac{ \sigma(\om,\m_1) }
		                          {(1+\fa(\om, \m_1))} \epc
\end{equation}
where $\sigma(\la,\m_1):=\frac{1}{\bar{\fa }(\la, \m_1)}
\frac{\partial}{\partial \delta} \bar{\fa }(\la, \m_1)$. One compares
(\ref{G}) with  (\ref{nliedelta}) and concludes
\begin{equation}\label{sigmaGrelation}
     \sigma(\la,\m_1) = 2 J \sh (\h) G(\la,\m_1;0) \epp
\end{equation}
Similarly we take the derivative of $\ln \Lambda(\mu_2, \m_1)$  with
respect to $\delta$ and find
\begin{align}
     \frac{\partial}{\partial \delta}\ln \Lambda(\m_2, \m_1) =&
     -\int_{\cal C} \frac{\rd \om}{2 \p \i} \,
     e(\om-\m_2) \frac{\s(\om,\m_1)}{(1+\fa(\om, \m_1))}
     = -\int_{\cal C} \frac{\rd \om}{2 \p \i} \,
     e(\om-\m_2) \frac{2 J \sh(\h) G(\om,\mu_1;0)}{(1+\fa(\om, \m_1))}
     \notag \\
     &=  - J \sh(\h) \int_{\cal C} \frac{\rd \om}{\p \i} \, 
     \frac{G(\om,\mu_1;0)}{(1+\fa(\om, \m_1))}
     \Bigl(    \coth(\om-\m_2-\h)  -\coth(\om-\m_2)    \Bigr) \epc
\end{align}
where we have used  (\ref{sigmaGrelation}) in the second equality.
By comparing the above equation with (\ref{psi}), one obtains
\begin{equation}
     \psi(\m_1,\mu_2;0) =\frac{1}{J \sh(\h)}
     \frac{\partial}{\partial \delta}\ln \Lambda(\m_2, \m_1)|_{\delta=0}
     \epp
\end{equation}
Then the evenness of $\psi(\m_1,\mu_2;0)$ follows from the same property
of the generalized transfer matrix.

Finally we show that $ \partial_{\al} \psi(\la_1,\la_2;\al)|_{\al=0}$
is also even. To prove this we consider the relation (\ref{dpmmp})
in appendix \ref{app:doubleint}. The lhs, $D^{+-}_{-+}(\la_1, \la_2)
+ D^{-+}_{+-}(\la_1, \la_2)$, is invariant  under $+\leftrightarrow -$,
hence it is even with respect to $h$. The first term in the rhs is
also even as it is proportional to $\psi(\mu_1,\mu_2;0)$ (see
(\ref {dqpsi})). Thus, the content of the bracket in the second term
of the rhs should be also even. Thanks to (\ref{onepointapp}) and
(\ref {dpp2}) it is represented as 
\[
     D^{+}_{+}(\la_1) + D^{+}_{+}(\la_2)-2 D^{++}_{++}(\la_1,\la_2) =
        \frac{\coth (\h)}{2} \psi(\la_1,\la_2;0)
	+ \frac{\coth (\la_1-\la_2)}{2\h }
	  \partial_{\al} \psi(\la_1,\la_2;\al) |_{\al=0} \epp
\]
Thus, we conclude that $ \partial_{\al} \psi(\la_1,\la_2;\al)|_{\al=0}$
is even.

\section{Thermal correlation functions of local operators}
\label{sec:corfactor}
In this section we are formulating our main result which is a
conjectured explicit formula for the correlation functions of
local operators in the XXZ chain at finite temperature and
finite magnetic field. The sources of this conjecture are the
results of the previous two sections that followed from the
finite temperature algebraic Bethe ansatz approach of \cite{GKS04a,%
GKS05,DGHK07} and the results of \cite{BJMST04b,BJMST06,BJMST06b},
where the exponential formula was discovered as a consequence of studying
the rqKZ equation. Unfortunately, both approaches differ considerably
in spirit and notation. We will try to reconcile them while keeping
as much as possible of the original notation. We have to ask the
reader to be forbearing though if this sometimes leads to confusion.

In \cite{BJMST06b} much emphasis was laid on developing a formalism
which applies directly to the infinite chain with lattice sites
$j \in {\mathbb Z}$. To keep things closely parallel we therefore
concentrate in this section on the temperature case and comment
on the finite length case only later in section~\ref{sec:concl}.
All operators $\CO$ which act non-trivially on any finite number
of lattice sites span a vector space~$\CW$. Because of the translational
invariance of the Hamiltonian we may content ourselves (as long as
we keep $\a = 0$) with operators which act non-trivially only on positive
lattice sites, $j \in {\mathbb N}$. We shall denote the restriction of
$\CO$ to the first $n$ lattice sites by $\CO_{[1,n]}$. The inhomogeneous
density matrix satisfies the reduction identity
\begin{equation} \label{reducedensmat}
     \tr_n D_n (\la_1, \dots, \la_n|T,h;0)
        = D_{n-1} (\la_1, \dots, \la_{n-1}|T,h;0) \epp
\end{equation}
It follows that the inductive limit
\begin{equation}
     \lim_{n \rightarrow \infty}
        \tr_{1, \dots, n} \bigl( D_n (\la_1, \dots, \la_n|T,h;0)
	                         \CO_{[1,n]} \bigr)
\end{equation}
exists and defines an operator $D^\ast_{T,h}: \CW \rightarrow
{\mathbb C}$ such that
\begin{equation}
     D^\ast_{T,h} (\CO) = \< \CO \>_{T,h}
\end{equation}
is the thermal average at finite magnetic field of the local operator
$\CO$ in the inhomogeneous XXZ model. Note that
\begin{equation}
     D^\ast_{T,h} \bigl( {e_1}^{\a_1}_{\be_1} \dots {e_n}^{\a_n}_{\be_n}
                  \bigr) =
        {D_n\,}_{\a_1, \dots, \a_n}^{\be_1, \dots, \be_n}
	      (\la_1, \dots, \la_n |T, h; 0) \epp
\end{equation}
For this reason we may interpret $D^\ast_{T,h}$ as a kind of `universal
density matrix' of the XXZ chain.

Let us define a linear functional $\mathbf{tr}: \CW \rightarrow
{\mathbb C}$ by
\begin{equation} \label{tracefun}
     \mathbf{tr}(\CO) = \dots \2 \tr_1\ \2 \tr_2\ \2 \tr_3 \dots (\CO)
                        \epp
\end{equation}
with $\tr_j$ the usual traces of $2 \times 2$ matrices. Then we
conjecture that an operator $\Om: \CW \rightarrow \CW$ exists such
that $D^\ast_{T,h} = \mathbf{tr} \; \re^\Om$. More precisely we propose
the following
\begin{conjecture*}
For all $\CO \in \CW$ the density matrix $D^\ast_{T,h}$ can be expressed
as
\begin{equation} \label{main2}
     D^\ast_{T,h} (\CO) = \mathbf{tr} \bigl(\re^{\Om} (\CO)\bigr) \epc
\end{equation}
where $\mathbf{tr}$ is the trace functional (\ref{tracefun}) and
$\Om: \CW \rightarrow \CW$ is a linear operator that can be decomposed
as
\begin{equation} \label{Omega12}
     \Om = \Om_1 + \Om_2
\end{equation}
with
\begin{subequations}
\label{Omega1and2}
\begin{align} \label{Omega1}
     \Om_1 & = - \lim_{\a \rightarrow 0}
                 \int \int \frac{\rd \m_1}{2 \p \i}
                 \frac{\rd \m_2}{2 \p \i} \:
		 \om(\m_1,\m_2;\a) \bb (\z _1;\a-1) \bc (\z _2;\a)
		 \epc \\[1ex]
     \Om_2 & = - \lim_{\a \rightarrow 0}
                 \int \frac{\rd \m_1}{2 \p \i}
                 \: \ph(\m_1;\a) \mathbf{H} (\z_1; \a)
		 \epp \label{Omega2}
\end{align}
Here $\z_j = \re^{\m_j}$, $j = 1, 2$, and $\om(\m_1, \m_2; \a)$ and
$\ph (\m_1; \a)$ are the functions defined in (\ref{om}) and
(\ref{phi}). The operators $\bb$, $\bc$ and $\mathbf{H}$ do not
depend on $T$ or $h$. They are purely algebraic. Their construction
will be explained below. The integrals mean to take residues at
the simple poles of $\bb$, $\bc$ and $\mathbf{H}$ located at the
inhomogeneities $\x_j$ (see below).
\end{subequations}
\end{conjecture*}

In fact, the operators $\bb$ and $\bc$ are the same as in the ground
state case \cite{BJMST06b}. The operator $\mathbf{H}$ is new in the
present context\footnote{Compare, however, eq.\ (\ref{Ha}) with
the operator $\mathbf{k}^{(0)}$ defined in Lemma A.2 of
\cite{BJMST07app}.}, but can be defined using the same algebraic notions
underlying the construction of $\bb$ and $\bc$. Note that
$\lim_{h \rightarrow 0} \ph(\m;0) = 0$ which implies that
$\lim_{h \rightarrow 0} \Om_2 = 0$. Hence, as
in the isotropic case \cite{BGKS06}, we observe that the algebraic
structure of the factorized form of the correlation functions is
identical in the ground state and for finite temperature as long as
the magnetic field vanishes. Due to the properties of the function
$\om$ we recover the result of \cite{BJMST06b} in the zero temperature
limit at vanishing magnetic field. In the high-temperature limit,
on the other hand, we conclude with (\ref{omom'beta}), (\ref{phibeta})
that $\lim_{T \rightarrow \infty} \Om = 0$ and that all correlation
functions trivialize in the expected way,
\begin{equation} \label{denshtc}
     \lim_{T \rightarrow \infty} D^\ast_{T,h} = \mathbf{tr} \epp
\end{equation}

For the definition of the operators $\bb$, $\bc$ and $\mathbf{H}$
we first of all generalize the space of local operators $\CW$ to
a space of quasi-local operators of the form
\begin{equation}
     \re^{\a \h \sum_{k = - \infty}^0 \s_k^z} \CO \epc
\end{equation}
where $\CO$ is local, and denote this space by $\CW_\a$. The operators
$\bb$, $\bc$ and $\mathbf{H}$ then act as
\begin{equation}
     \bb (\z; \a): \CW_\a \rightarrow \CW_{\a + 1} \epc \qd
     \bc (\z; \a): \CW_\a \rightarrow \CW_{\a - 1} \epc \qd
     \mathbf{H} (\z; \a): \CW_\a \rightarrow \CW_\a
\end{equation}
which implies in particular that $\bb (\z _1;\a-1) \bc (\z _2;\a):
\CW_\a \rightarrow \CW_\a$.

The $z$-component of the total spin is the formal series
$S_\infty^z$ (see equation (\ref{totalsz})). We denote its adjoint
action by
\begin{equation}
     {\mathbb S} (X) = [S^z_\infty, X] \epp
\end{equation}
Then $q^{\a \mathbb S}: \CW_\a \rightarrow \CW_\a$. The spin reversal
operator defined by
\begin{equation}
     {\mathbb J} (X) = \Bigl[ \prod_{j \in {\mathbb Z}} \s_j^x \Bigr] X
                       \Bigl[ \prod_{j \in {\mathbb Z}} \s_j^x \Bigr]
\end{equation}
clearly is a map ${\mathbb J}: \CW_\a \rightarrow \CW_{- \a}$.

The operators $\bb$, $\bc$ and $\mathbf{H}$ will be defined in two
steps. We first define endomorphisms $\bb_{[kl]}$, $\bc_{[kl]}$ and
$\mathbf{H}_{[kl]}$ acting on $\End (\CV)$, where the tensor product
$\CV = V_k \otimes \dots \otimes V_l$ represents the space of states of
a segment of the infinite spin chain reaching from site $k$ to site $l$,
and $V_j$ is isomorphic to ${\mathbb C}^2$. Then we use that these
endomorphisms have a reduction property similar to (\ref{reducedensmat})
which allows us to extend their action to $\CW_\a$ by an inductive
limit procedure. The endomorphisms $\bb_{[kl]}$, $\bc_{[kl]}$ and
$\mathbf{H}_{[kl]}$ are constructed from weighted traces of the
elements of certain monodromy matrices related to $U_q
(\widehat{\mathfrak{sl}}_2)$. These monodromy matrices are obtained
from products of $L$-matrices with different auxiliary spaces.

The simplest case is directly related to the $R$-matrix of the six-vertex
model,
\begin{align}
     & R(\zeta) = (q\z-q^{-1}\z^{-1})
                  \begin{pmatrix}
		     1&0&0&0\\
		     0&\be(\z)&\g(\z)&0\\
		     0&\g(\z)&\be(\z)&0\\
		     0&0&0&1
		  \end{pmatrix} \epc
\end{align}
where
\begin{equation}
     \be(\z) = \frac{(1 - \z^2)q}{1 - q^2 \z^2} \epc \qd
     \g(\z) = \frac{(1 - q^2)\z}{1 - q^2 \z^2}
\end{equation}
and $q = \re^{\a \h}$. Let us fix an auxiliary space $V_a$
isomorphic to ${\mathbb C}^2$. Then $L_{a,j} (\z) = R_{a,j} (\z)$
is the standard $L$-matrix of the six-vertex model. The corresponding
monodromy matrix is
\begin{equation}
     T_{a,[k,l]} (\z) = L_{a,k} (\z/\x_k) \dots L_{a,l} (\z/\x_l) \epp
\end{equation}
It acts on $V_a \otimes \CV$. We are interested in operators acting
on $\End(\CV)$. Such type of operators are naturally given by the
adjoint action of operators acting on $\CV$. An example is the
transfer matrix $\mathbf{t}_{[k,l]} (\z)$ defined by
\begin{equation}
     \mathbf{t}_{[k,l]} (\z) (X) = \tr_a T_{a,[k,l]} (\z)^{-1} X
                                         T_{a,[k,l]} (\z)
\end{equation}
for all $X \in \End(\CV)$. It will be needed in the definition of the
operator $\mathbf{H}_{[k,l]}$ below.

Further following \cite{BJMST06b} we introduce another type of 
monodromy matrices for which the auxiliary space is replaced with
the $q$-oscillator algebra $Osc$ generated by $a,a^*,q^{\pm D}$
modulo the relations
\begin{align} \label{alg}
     & q^D a^\ast = a^\ast q^{D+1} \epc &
     & q^D a = a q^{D-1} \epc \notag \\
     & a^\ast a = 1 - q^{2D} \epc &
     & a a^\ast = 1 - q^{2D+2} \epp
\end{align}
We consider two irreducible modules $W^\pm$ of $Osc$,
\begin{equation} \label{rep+}
     W^+ = \bigoplus\limits_{k \ge 0} \mathbb{C}|k\> \epc \qd
     W^- = \bigoplus\limits_{k \le - 1} \mathbb{C}|k\> \epc
\end{equation}
defined by the action
\begin{equation}
     q^D |k\> = q^k |k\> \epc \qd
     a |k\> = (1 - q^{2k})|k - 1\> \epc \qd
     a^\ast |k\> = (1 - \de_{k, -1})|k + 1\>
\end{equation}
of the generators. The $L$-operators $L^\pm (\z) \in Osc \otimes
\End (\CV)$ are defined by
\begin{subequations}
\begin{align} \label{L+}
     L^+ (\z) & = \i \z^{-1/2} q^{-1/4} (1 - \z a^\ast \s^+
                     - \z a \s^- - \z^2 q^{2D+2} \s^- \s^+) q^{\s^z D}
		     \epc \\[1ex]
     L^- (\z) & = \s^x L^+ (\z) \s^x \epp
\end{align}
\end{subequations}
The corresponding monodromy matrices are
\begin{equation} \label{monodromy}
     T^\pm_{A,[k,l]}(\z) = L^\pm_{A,l} (\z/\x_l) \dots
                           L^\pm_{A,k} (\z/\x_k) \epc
\end{equation}
where the index $A$ refers to the auxiliary space $Osc$. We denote
their (inverse) adjoint action by
\begin{equation}
     \mathbb{T}^\pm_{A,[k,l]} (\z)^{-1} (X) =
        T^\pm_{A,[k,l]} (\z)^{-1} X T^\pm_{A,[k,l]}(\z)
\end{equation}
for all $X \in \End (\CV)$. Here the inverse on the right hand side is
taken for both auxiliary and `quantum' space. The analogue of the transfer
matrix $\mathbf{t}_{[k,l]}$ in this case are two $Q$-operators $Q^\pm$
(see \cite{BJMST06b}). Since we need only one of them here we leave
out the superscript and define\footnote{Here we use a slightly
different definition of $Q$-operator in comparison with
$\mathbf{Q}^+$ in \cite{BJMST06b}, see formula (2.10) there. The
difference is an additional factor $(1-q^{2(\a - {\mathbb S})})$.}
\begin{equation} \label{Qop}
     \mathbf{Q}_{[k,l]} (\z,\a) = \tr^+_A
        \bigl( q^{2\al D_A} \mathbb{T}^+_{A,[k,l]}(\z)^{-1} \bigr) \epp
\end{equation}
Here $\tr^+_A$ signifies that the trace is taken over $W^+$. Similarly
we will denote the trace over $W^-$ by $\tr^-_A$.

Now we are prepared to define the restriction of the operator
$\mathbf{H}$ to $\End(\CV)$,
\begin{equation} \label{H}
     \mathbf{H}_{[k,l]} (\z;\al) = \mathbf{Q}_{[k,l]} (\z;\al)
                                   \mathbf{t}_{[k,l]} (\z) \epp
\end{equation}
We show below that this definition (in the limit $\a \rightarrow 0$)
can be inductively extended to $\CW_\a$. To avoid possible confusion let
us note that in fact the operator $\mathbf{H}$ defined by the formula
(\ref{H}) {\it is not the left hand side of Baxter's $TQ$-relation.} In
order that it were we would need to `$\a$-deform' the $\mathbf{t}$-operator
as well.

In order to obtain $\bb_{[k,l]}$ and $\bc_{[k,l]}$ and also another
form of the operator $\mathbf{H}_{[k,l]}$ we recall the fusion
technique used in \cite{BJMST06b}. There the fused $L$-operators
\begin{equation}
     L^\pm_{\{A,a\}, j} (\z) =
        (G^\pm_{A,a})^{-1}  L^\pm_{A,j} (\z) \, R_{a,j}(\z) G^\pm_{A,a}
\end{equation}
were defined, where
\begin{equation}
     G^\pm_{A,a} = q^{\mp \s_a^z D_A} (1 + a^\ast_a \s^\pm) \epp
\end{equation}
The application of $G^+_{A,a}$ transforms $L^\pm_{A,j} (\z) R_{a,j}(\z)$
into a matrix of lower triangular form on $V_a$,
\begin{equation} \label{fusion}
     L^+_{\{A,a\}, j} (\z) = (\z q-\z ^{-1}q^{-1})
         \begin{pmatrix}
	    L^+_{A,j} (q^{-1} \z) q^{- \s^z_j/ 2} & 0 \\
	    \g (\z) L^+_{A,j}(q \z) \s_j^+ q^{-2 D_A + 1/2} &
	    \be(\z) L^+_{A,j} (q \z) q^{\s^z_j/ 2}
	 \end{pmatrix}_a \epp
\end{equation}
The inverse is also of lower triangular form and is given by
\begin{multline} \label{fusioninv}
     L ^+_{\{A,a\},j} (\z)^{-1} = \frac 1{q \z - q^{-1} \z^{-1}}
          \\ \times
	  \begin{pmatrix}
             q^{\s_j^z/2} L^+_{A,j}(q^{-1} \z)^{-1} & 0 \\
	     - \g(q^{-1} \z) \s_j^+ q^{-2D_A - 1/2}
	       L^+_{A,j} (q^{-1} \z)^{-1}
	     & \be(\z)^{-1} q^{- \s^z_j/2} L^+_{A,j} (q \z)^{-1}
	  \end{pmatrix}_a \epp
\end{multline}
Correspondingly
\begin{equation}
     L ^-_{\{A,a\},j} (\z) =
        \s_a^x \s_j^x \, L ^+_{\{A,a\},j} (\z) \, \s_a^x \s_j^x
\end{equation}
is of upper triangular form. It follows that similar statements hold
for the monodromy matrices
\begin{equation}
     T^\pm_{\{A,a\},[k,l]} (\z) =
        (G^\pm_{A,a})^{-1} T^\pm_{A,[k,l]} (\z) T_{a,[k,l]} (\z)
	G_{A,a}^\pm \epp
\end{equation}
$T^+_{\{A,a\},[k,l]} (\z)$ acts as a lower triangular matrix in
$V_a$, $T^-_{\{A,a\},[k,l]} (\z)$ as an upper triangular matrix.
As before we are interested in the adjoint action of the fused
monodromy matrices on endomorphisms $X \in \End(\CV)$. Following
\cite{BJMST06b} we define
\begin{equation}
     \mathbb{T}^\pm_{\{A,a\},[k,l]} (\z)^{-1} (X) =
        T^\pm_{\{A,a\},[k,l]} (\z)^{-1} X T^\pm_{\{A,a\},[k,l]}(\z)
\end{equation}
for all $X \in \End(\CV)$.

Regarding $\mathbb{T}^\pm_{\{A,a\},[k,l]} (\z)^{-1}$ as matrices acting
on $V_a$ as in \cite{BJMST06b} we may write their entries as
\begin{align}
     \bT^{+}_{\{A,a\},[k,l]} (\z)^{-1} & =
        \begin{pmatrix}
           \bbA^+_{A,[k,l]} (\z) & 0 \\
	   \bbC^+_{A,[k,l]} (\z) & \bbD^+_{A,[k,l]} (\z)
        \end{pmatrix}_a \epc \notag \\[1ex]
     \bT^{-}_{\{A,a\},[k,l]} (\z)^{-1} & =
        \begin{pmatrix}
           \bbA^-_{A,[k,l]} (\z) & \bbB^-_{A,[k,l]} (\z) \\
	   0 & \bbD^-_{A,[k,l]} (\z)
        \end{pmatrix}_a \epp
\end{align}
The entries of these matrices are elements of $Osc \otimes \End(\CV)$.
We are now prepared to define $\bb_{[k,l]}$ and $\bc_{[k,l]}$,
\begin{subequations}
\begin{align} \label{defc}
     \bc_{[k,l]} (\z, \a) & =
         q^{\a - {\mathbb S}_{[k,l]}} (1 - q^{2(\a - {\mathbb S}_{[k,l]})})
         \sing \bigl[ \z^{\a - {\mathbb S}_{[k,l]}}
               \tr_A^+ \bigl( q^{2 \a D_A}
                              {\mathbb C}^+_{A, [k,l]} (\z) \bigr) \bigr]
               \epc \\[1ex] \label{defb}
     \bb_{[k,l]} (\z, \a) & =
         q^{2 {\mathbb S}_{[k,l]}}
         \sing \bigl[ \z^{- \a + {\mathbb S}_{[k,l]}}
               \tr_A^- \bigl( q^{- 2 \a (D_A + 1)}
                              {\mathbb B}^-_{A, [k,l]} (\z) \bigr)
               \bigr] \epp
\end{align}
\end{subequations}
The symbol `sing' means taking the singular part at $\z = \x_j$, $j =
1, \dots, n$ (cf.\ eq.\ (2.13) of \cite{BJMST06b}). These operators
raise or lower the $z$-component of the total spin by one,
\begin{equation}
     [{\mathbb S}_{[k,l]}, \bc_{[k,l]} (\z, \a)] = \bc_{[k,l]} (\z, \a)
        \epc \qd
     [{\mathbb S}_{[k,l]}, \bb_{[k,l]} (\z, \a)] = - \bb_{[k,l]} (\z, \a)
        \epp
\end{equation}
Their properties were extensively studied in \cite{BJMST06b,BJMST07app}.
Here we shall only need the following.
\begin{proposition}
Reduction properties \cite{BJMST06b}.
\begin{align} \label{redpropcb}
     \bc_{[k,l]} (\z, \a) \bigl(X_{[k,l-1]} \, I_l\bigr) & =
        \bc_{[k,l-1]} (\z, \a) \bigl(X_{[k,l-1]}\bigr) \, I_l \epc
        \notag \\[1ex]
     \bb_{[k,l]} (\z, \a) \bigl(X_{[k,l-1]} \, I_l\bigr) & =
        \bb_{[k,l-1]} (\z, \a) \bigl(X_{[k,l-1]}\bigr) \, I_l \epc
        \notag \\[1ex]
     \bc_{[k,l]} (\z, \a) \bigl(q^{\a \s^z_k} \, X_{[k+1,l]}\bigr) & =
        q^{(\a - 1) \s^z_k} \, \bc_{[k+1,l]} (\z, \a)
        \bigl(X_{[k+1,l]}\bigr) \epc \notag \\[1ex]
     \bb_{[k,l]} (\z, \a) \bigl(q^{\a \s^z_k} \, X_{[k+1,l]}\bigr) & =
        q^{(\a + 1) \s^z_k} \, \bb_{[k+1,l]} (\z, \a)
        \bigl(X_{[k+1,l]}\bigr) \epp
\end{align}
\end{proposition}
From this it follows that $\bc_{[k,l]} (\z, \a)$ can be inductively
extended to an operator $\bc (\z, \a): \CW_\a \rightarrow \CW_{\a-1}$.
Similarly $\bb_{[k,l]} (\z, \a)$ inductively extends to an operator
$\bb (\z, \a): \CW_\a \rightarrow \CW_{\a+1}$. These are the operators
appearing in the definition (\ref{Omega1}) of $\Om_1$.

Using the simple relation
\begin{equation}
     (G^+_{A,a})^{-1} q^{2 \a D_A} G^+_{A,a} =
        q^{2\al D_A} \begin{pmatrix}
	                1 & (1-q^{-2\al}) a^{\ast}_A\\
			0 & \quad\quad 1
                     \end{pmatrix}_a
\end{equation}
and the concrete form of $L ^+_{\{A,a\},j}(\z )$ and
$L ^+_{\{A,a\},j}(\z )^{-1}$ one can obtain
\begin{equation} \label{Ha}
     \mathbf{H}_{[k,l]} (\z; \a) \backsimeq
        (1 - q^{-2\a}) \tr_A^+ \bigl( q^{2 \a D_A} a^{\ast}_A
	                              \bbC^+_{A,[k,l]} (\z) \bigr) \epc
\end{equation}
where the symbol $\backsimeq$ means equality up to the regular part
when $\z\rightarrow \x_j$. Since the function $\ph (\mu,\al)$ is regular
when $\mu\rightarrow 0$, the regular part of $\mathbf{H}_{[k,l]}(\z;\a)$
does not contribute to the right hand side of (\ref{Omega2}). The formula
(\ref{Ha}) looks rather similar to the definition (\ref{defc}) of the
operator $\bc_{[k,l]}$. The essential difference is due to the insertion
of $a^{\ast}_A$ under the trace. In contrast to the $\bc_{[k,l]}$-operator
which increases the total spin, the operator $\mathbf{H}_{[k,l]}$ does
not change the total spin.

\subsection*{Properties of the operators $\Omega_1$ and $\Omega_2$}
Assuming for a moment that the limit on the right hand side of
(\ref{Omega1}) exists we can conclude with (\ref{redpropcb}) that
\begin{align} \label{redOm1}
     & (\Om_1)_{[k,l]} (X_{[k,l-1]} \, I_l) =
        (\Om_1)_{[k,l-1]} \, (X_{[k,l-1]}) \, I_l \epc \notag \\[1ex]
     & (\Om_1)_{[k,l]} (I_k \, X_{[k+1,l]}) =
        I_k \, (\Om_1)_{[k+1,l]} (X_{[k+1,l]}) \epp
\end{align}
Due to this property one can define $\Om_1$ as the inductive limit of
its restriction
\begin{equation} \label{limOm1}
     \Om_1 = \lim_{k \rightarrow - \infty} \lim_{l \rightarrow \infty}
             (\Om_1)_{[k,l]} \epp
\end{equation}
As we shall discuss later the same is also true for the operator $\Om_2$.

But before we come to this point let us check whether the limits in
the right hand side of (\ref{Omega1}) and (\ref{Omega2}) are really
well defined. 
\begin{proposition}
The limits in the right hand side of eqs.\ (\ref{Omega1}) and
(\ref{Omega2}) exist.
\end{proposition}
\begin{proof}
The existence of the limit in (\ref{Omega2}) follows from the
formula (\ref{Ha}), because taking the trace there can results in
at most a simple pole $1/(1-q^{\al})$. This pole will be canceled
by the factor $(1-q^{-2\al})$ which stands in front of the trace 
in (\ref{Ha}).

In order to prove the existence of the limit in (\ref{Omega1}) we use
an alternative representation of~$\Om_1$,
\begin{equation} \label{Omega1a}
     \Om_1 = - \lim_{\a \rightarrow 0} \left[ \frac1{q^\a - q^{- \a}}
               \int \int \frac{\rd \m_1}{2 \p \i} \frac{\rd \m_2}{2 \p \i}
               \left( \frac{\z_1}{\z_2} \right)^\a
               \om (\m_2, \m_1; \a) \widetilde{\bX} (\z_1,\z _2;\a)
               \right] \epc
\end{equation}
where\footnote{Here we take only the spin-0 sector.}
\begin{equation} \label{X}
     \widetilde{\bX} (\z _1,\z _2;\a) = \sing_{\z_1, \z_2}
        \bigl[ \tr _{a,b}\( B_{a,b} (\z_1/\z_2)
               \bT_b (\z_2)^{-1} \bT_a(\z _1)^{-1}\)
               \bQ^-(\z_2;\al)\bQ^+(\z_1;\al )\bigr]
\end{equation}
with the `boundary' matrix
\begin{equation}
     B (\z) = \frac{(\z -\z ^{-1})}
                   {2(\z q -\z ^{-1}q^{-1})(\z q^{-1} -\z ^{-1}q)}
              \begin{pmatrix}
                  0 &0 &0 &0\\
                  0 &\z +\z ^{-1} & -q-q^{-1} &0\\
                  0  & -q-q^{-1} &  \z +\z ^{-1} &0\\
                  0 &0 &0 &0
              \end{pmatrix}
\end{equation}
and $\bQ^{\pm}$ the same operators as defined in \cite{BJMST06b},
\begin{subequations}
\begin{align}
     & \mathbf{Q}^+_{[1,n]} (\z, \a) =
       \tr^+_A \(q^{2 \a D_A}\ \bT^{+}_{A,[1,n]}(\z)^{-1}\)
                (1-q^{2(\a - {\mathbb S})}) \epc \label{Q+} \\[1ex]
     & \mathbf{Q}^-_{[1,n]} (\z, \a) =
       \tr^-_A\(q^{-2 \a (D_A+1)}\ \bT ^{-}_{A,[1,n]}(\z)^{-1}\)
                q^{2 {\mathbb S}} (1-q^{2(\a - {\mathbb S})})
                \epp \label{Q-}
\end{align}
\end{subequations}
The form (\ref{Omega1a}) of $\Om_1$ is similar to the form shown in the
appendix of \cite{BJMST06b}. It can be obtained combining the ideas
of \cite{BJMST06} and \cite{BJMST06b}.

The limit in (\ref{Omega1a}) exists, since the integrand is
antisymmetric in $\z_1$, $\z_2$ in the limit $\a \rightarrow 0$. This
can be seen as follows. First of all $\om (\m_2, \m_1; \a) (\z_1/\z_2)^\a$
is symmetric in $\z_1$, $\z_2$ for $\a \rightarrow 0$ (see eq.\
(\ref{omexchange1})). Next $\tr _{a,b}\( B_{a,b} (\z_1/\z_2)
\bT_b (\z_2)^{-1} \bT_a(\z _1)^{-1}\)$ is independent of $\a$ and
antisymmetric in $\z_1$, $\z_2$, since $B(\z_1/\z_2)$ is antisymmetric
in $\z_1$, $\z_2$ and since $[B(\z_1), R(\z_2)] = 0$.

It remains to show that $\bQ^-(\z_2;\al)\bQ^+(\z_1;\al )$ is symmetric
for $\a \rightarrow 0$. This product is meromorphic in $\a$ by
construction. We show by an explicit calcualtion in appendix
\ref{app:proofs} that it is regular at $\a = 0$ and symmetric in $\z_1$,
$\z_2$ in this point. In fact, adopting the notation
\begin{equation} \label{basis}
     \bQ^{\pm}(\z;0) \bigl( {e_1}_{\s_1}^{\e_1} \dots
                            {e_n}_{\s_n}^{\e_n} \bigr) =
     \sum_{\s'_1,\dots,\s'_n;\e'_1,\dots,\e'_n}
         \bigl[ \bQ^{\pm}(\z;0) \bigl]^%
                {\s_1, \dots, \s_n;\;\e_1, \dots,\e_n}_%
                {\s'_1, \dots, \s'_n;\;\e'_1, \dots,\e'_n}\
                {e_1}_{\s'_1}^{\e'_1} \dots {e_n}_{\s'_n}^{\e'_n}
\end{equation}
for the matrix elements of the operators $\bQ^{\pm}(\z;0)$ with respect
to the canonical basis we obtain
\begin{multline} \label{Qexplicitinhom}
     \bigl[ \bQ^{\pm} (\z; 0) \bigl]^%
            {\s_1, \dots, \s_n; \;\e_1, \dots, \e_n}_%
            {\s'_1,\dots, \s'_n; \;\e'_1, \dots, \e'_n} =
     \delta_{\e_1 + \dots + \e_n, \s_1 + \dots + \s_n}
     \delta_{\e'_1 + \dots + \e'_n, \s'_1 + \dots + \s'_n} \\
     \times \Biggl[ \prod_{j=1}^n
        \frac{\e_j \e'_j (\z/\xi_j)^{- \frac12 (\e_j \e'_j + \s_j \s'_j)}}
             {\z/\xi_j - \xi_j/\z} \Biggr]
       q^{\frac12 \sum_{1 \le j < k \le n}
          \bigl((\e_j - \e'_j)\e'_k - (\s_j - \s'_j)\s'_k \bigr)} \epp
\end{multline}
Hence,
\begin{multline}
     \bQ^-(\z_2;0) \bQ^+(\z_1;0) \\ = \bQ^+(\z_2;0) \bQ^+(\z_1;0)
        = \bQ^+(\z_1;0) \bQ^+(\z_2;0) = \bQ^-(\z_1;0) \bQ^+(\z_2;0) \epc
\end{multline}
where we used the commutativity $[\bQ^+(\z_1; \a), \bQ^+(\z_2; \a)] = 0$
(see \cite{BJMST07app}) in the second equation.
\end{proof}

Following the same lines one can show that the operator $\Om_1$ is
symmetric under the spin reversal transformation,
\begin{equation} \label{spinrevOm1}
     \Om_1 = \bJ \Om_1 \bJ \epp
\end{equation}
Moreover, $\Om_1$ is symmetric under reversal of the direction of the
magnetic field
\begin{equation} \label{hOm1}
     \Om_1 = \Om_1 \bigr|_{h \leftrightarrow -h} \epc
\end{equation}
since $\om$ is an even function of the magnetic field $h$. An actual
calculation of the right hand side of eq.\ (\ref{Omega1}) or
(\ref{Omega1a}) demands to apply l'H\^opital's rule. As a result one
gets two terms: one standing with $\om(\mu_1,\mu_2;0)$ which is even
with respect to the transposition of $\mu_1$ and $\mu_2$ and another
one with $\om'(\mu_1,\mu_2;0)$ which is odd with respect to
$\mu_1\leftrightarrow\mu_2$. This is the same splitting as discussed
in the paper \cite{BJMST04b}. Below in section~\ref{sec:examples} we
will consider several examples in order to illustrate this point.

Let us now come to the properties of the operator $\Om_2$. We shall
consider
\begin{equation} \label{Hj}
     \mathbf{H}_j
         \bigl({e_1}_{\s_1}^{\e_1} \dots {e_n}_{\s_n}^{\e_n} \bigr) =
     \lim_{\a \rightarrow 0} \res_{\z = \x_j}
     \mathbf{H}_{[1,n]}
         \bigl({e_1}_{\s_1}^{\e_1} \dots {e_n}_{\s_n}^{\e_n}\bigr) \epp
\end{equation}
In the following we shall need an explicit formula which is also proved in
appendix \ref{app:proofs},
\begin{equation} \label{H1explicit}
     \mathbf{H}_1
        \bigl({e_1}_{\s_1}^{\e_1} \dots {e_n}_{\s_n}^{\e_n}\bigr) =
        \bigl( \bQ_{\s_1}^{\e_1} \bR_{1;2,\cdots,n} \bigr)
        \bigl({e_2}_{\s_2}^{\e_2} \dots {e_n}_{\s_n}^{\e_n} \bigr) \epc
\end{equation}
where the action of the operator $\bR_{1;2,\cdots,n}$ is defined by
\begin{equation}
     \bR_{1;2,\cdots,n} \bigl( X_{[2,n]} \bigr) =
        R_{2,1} \dots R_{n,1} X_{[2,n]} R_{1,n} \dots R_{1,2}
\end{equation}
with the standard $R$-matrix of the six-vertex model $R_{i,j} =
R_{i,j} (\xi_i/\xi_j)$ and where the matrix elements of the operator
$\bQ_{\s_1}^{\e_1}$ are explicitly given by
\begin{multline} \label{Q1explicit}
     \bigl[ \bQ_{\s_1}^{\e_1} \bigl]^%
        {\s_2, \dots, \s_n; \; \e_2, \dots, \e_n}_%
        {\s'_2, \dots, \s'_n; \; \e'_2, \dots, \e'_n} =
     \delta_{\e_1 + \dots + \e_n, \s_1 + \dots + \s_n}
     \delta_{\e'_2 + \dots + \e'_n, \s'_2 + \dots + \s'_n} \\
     \times \frac{\e_1}2 \prod_{j=2}^n
            \frac{\e_j \e'_j
            (\xi_j/\xi_1)^{\frac{\e_j \e'_j + \s_j \s'_j}2}}
                                {\x_j/\x_1 - \x_1/\x_j}
     q^{\frac12 \bigl\{ (\e_1 - \s_1)
        \sum_{2 \le k \le n}\e'_k +
        \sum_{2 \le j < k \le n} \bigl((\e_j - \e'_j) \e'_k
                                         - (\s_j - \s'_j) \s'_k\bigr)
        \bigr\}} \epp
\end{multline}
Note that the limit $\a \rightarrow 0$ and the calculation of the
residue at $\z = \x_1$ in equation (\ref{H1explicit}) may not be
interchanged.

The $\a = 0$ limit of the residues at $\z = \x_j$ for $j\ge 2$ can
be obtained from the formula (\ref{H1explicit}) by applying the exchange
relations
\begin{equation} \label{exchange}
     \check{\bR}_{i,i+1} \, \mathbf{H}_{[1,n]} (\z; \a)
        \bigl( X_{[1,n]} \bigr) = \mathbf{H}^{(i,i+1)}_{[1,n]} (\z; \a) \,
        \check{\bR}_{i,i+1} \bigl( X_{[1,n]} \bigr)
\end{equation}
with $\mathbf{H}^{(i,i+1)}_{[1,n]} = {\mathbf{H}_{[1,n]}|}_%
{\xi_i\leftrightarrow\xi_{i+1}}$ and the action 
\begin{equation}
     \check{\bR}_{i,i+1} \bigl( X_{[1,n]} \bigr) =
        \check{R}_{i,i+1} X_{[1,n]} \check{R}_{i,i+1}^{-1}
\end{equation}
for $1 \le i, i+1 \le n$. For example,
\begin{equation} \label{H2explicit}
     \mathbf{H}_2
       \bigl( {e_1}_{\s_1}^{\e_1} \dots {e_n}_{\s_n}^{\e_n} \bigr) =
           R_{\e_2,\e_1}^{\e''_2,\e''_1} (\x_2/\x_1)
           R_{\s''_1,\s''_2}^{\s_1,\s_2} (\x_1/\x_2)
          \bigl( \bQ_{\s''_2}^{\e''_2} \bR_{2;3,\cdots,n} \bigr)
          \bigl( {e_1}_{\s''_1}^{\e''_1} {e_3}_{\s_3}^{\e_3} \dots
                 {e_n}_{\s_n}^{\e_n} \bigr) \epp
\end{equation}

A most important consequence of the explicit formula (\ref{Q1explicit})
is the reduction property 
\begin{proposition}
\label{reducehs}
\begin{subequations}
\begin{align}
     & \mathbf{H}_1
       \bigl( I_1 \; {e_2}_{\s_2}^{\e_2} \dots {e_n}_{\s_n}^{\e_n}
              \bigr) = 0 \epc \label{redrightH1} \\[1ex]
     & \mathbf{H}_j
       \bigl (I_1 \; {e_2}_{\s_2}^{\e_2} \dots {e_n}_{\s_n}^{\e_n} \bigr) =
              I_1 \; \mathbf{H}_j \bigl({e_2}_{\s_2}^{\e_2} \dots
                                        {e_n}_{\s_n}^{\e_n} \bigr) \epc
       \qd 2 \le j \le n \label{redrightHj} \epc \\[1ex]
     & \mathbf{H}_j
       \bigl( {e_1}_{\s_1}^{\e_1} \dots {e_{n-1}}_{\s_{n-1}}^{\e_{n-1}} \;
              I_n \bigr) =
       \mathbf{H}_j
       \bigl( {e_1}_{\s_1}^{\e_1} \dots {e_{n-1}}_{\s_{n-1}}^{\e_{n-1}}
              \bigr) \; I_n \epc \qd 1 \le j \le n-1 \epc
       \label{redleftHj} \\[1ex]
     & \mathbf{H}_n
       \bigl( {e_1}_{\s_1}^{\e_1} \dots {e_{n-1}}_{\s_{n-1}}^{\e_{n-1}} \;
              I_n \bigr) = 0 \epp \label{redleftHn}
\end{align}
\end{subequations}
\end{proposition}
\begin{proof}
The first formula (\ref{redrightH1}) is rather trivial because 
from the formula (\ref{Q1explicit}) it follows that
\[
     \sum_{\s = \pm 1} \bigl[ \bQ_{\s}^{\s} \bigr]^%
        {\s_2, \dots, \s_n; \e_2, \dots, \e_n}_%
        {\s'_2, \dots, \s'_n; \e'_2, \dots, \e'_n} = 0 \epp
\]
The second formula (\ref{redrightHj}) is less trivial. Let us outline
the proof for $j=2$. First we use (\ref{H2explicit}) in order to obtain
\begin{equation} \label{H2id1explicit}
     \mathbf{H}_2
       \bigl( I_1 \; {e_2}_{\s_2}^{\e_2} \dots {e_n}_{\s_n}^{\e_n} \bigr)
         = R_{\e_2, \e_1}^{\e''_2,\e''_1} (\x_2/\x_1)
           R_{\s''_1,\s''_2}^{\e_1, \s_2} (\x_1/\x_2)
       \bigl( \bQ_{\s''_2}^{\e''_2} \; \bR_{2; 3, \dots, n} \bigr)
       \bigl( {e_1}_{\s''_1}^{\e''_1} {e_3}_{\s_3}^{\e_3} \dots
              {e_n}_{\s_n}^{\e_n} \bigr)
\end{equation}
and substitute eq.\ (\ref{Q1explicit}). The latter should be separated
into two parts in such a way that only one of them is touched by two
$R$-matrices in the right hand side of (\ref{H2id1explicit}). This part
looks like
\begin{multline*}
     V_{\s'_1,\s''_1,\s''_2}^{\e'_1,\e''_1,\e''_2} (\x_1/\x_2) \\ :=
        \frac12 \e'_1 \e''_1 \e''_2
        \frac{(\x_1/\x_2)^{\frac12 (\e'_1 \e''_1 + \s'_1 \s''_1)}}
             {\x_1/\x_2 - \x_2/\x_1} \cdot 
     q^{\frac12 \bigl( (\e''_2 - \s''_2) \e'_1 -
                       (\s''_1 - \s'_1)(\e'_1 - \s'_1)\bigr)} \cdot 
     q_1^{\frac12 \bigl(\e''_1 + \e''_2 - \s''_1 -
                        \s''_2 - \e'_1 + \s'_1 \bigr)} \epc
\end{multline*}
where $q_1=q^{\e'_3+\cdots+\e'_n}$ and where the indices $\e'_3, \dots,
\e'_n$ are considered to be fixed. The following identity can be verified 
directly, for example, on a computer
\[
     V_{\s'_1, \s''_1, \s''_2}^{\e'_1, \e''_1, \e''_2} (\x_1/\x_2) \;
        R_{\e_2, \e_1}^{\e''_2, \e''_1} (\x_2/\x_1)
        R_{\s''_1, \s''_2}^{\e_1, \s_2} (\x_1/\x_2) =
        \frac12 \de_{\s'_1,\e'_1} \e_2 \; q_1^{\frac12(\e_2-\s_2)} \epp
\]
If we substitute the right hand side back into (\ref{H2id1explicit}) and 
collect all pieces we come to the statement that we wanted to prove,
namely, 
\[
     \mathbf{H}_2 \bigl( I_1 \; {e_2}_{\s_2}^{\e_2} \dots
                         {e_n}_{\s_n}^{\e_n} \bigr) =
     I_1 \mathbf{H}_2 \bigl({e_2}_{\s_2}^{\e_2} \dots
                            {e_n}_{\s_n}^{\e_n} \bigr) \epp
\]
The other cases when $j>2$ can be treated in a similar way. The formulae
(\ref{redleftHj}), (\ref{redleftHn}) are simple consequences of the
inversion of $L$-operators in the definition (\ref{H}).
\end{proof}

Using proposition \ref{reducehs} one immediately comes to the reduction
relation for $\Om_2$ because the restriction of (\ref{Omega2}) to the
interval $[1,n]$ is
\begin{equation} \label{Omega2Hj}
     (\Om_2)_{[1,n]} = - \sum_{j=1}^n \ph (\la_j; 0) \mathbf{H}_j \epp
\end{equation}
\begin{proposition}
Reduction identity for $\Om_2$.
\begin{align} \label{redOm2}
     &(\Om_2)_{[1,n]} \( X_{[1,n-1]} \; I_n \) =
      (\Om_2)_{[1,n-1]} \( X_{[1,n-1]} \) \; I_n \epc \notag \\[1ex]
     &(\Om_2)_{[1,n]} \(I_1 \; X_{[2,n]} \) =
       I_1 \; (\Om_2)_{[2,n]} \( X_{[2,n]} \) \epp
\end{align}
Due to (\ref{redOm2}) we may define $\Om_2$  for the infinite chain
through an inductive limit as in eq.\ (\ref{limOm1}).
\end{proposition}

Another immediate consequence of the formula (\ref{Q1explicit}) is the
spin reversal anti-symmetry. First of all
\begin{equation} \label{spinrevQ1}
     \bigl[ \bQ_{-\s_1}^{-\e_1} \bigl]^%
            {-\s_2, \dots, -\s_n; -\e_2, \dots, - \e_n}_%
            {-\s'_2, \dots, -\s'_n; -\e'_2, \dots, -\e'_n} =
     - \bigl[ \bQ_{\s_1}^{\e_1} \bigl]^%
            {\s_2, \dots, \s_n; \e_2, \dots, \e_n}_%
            {\s'_2, \dots, \s'_n; \e'_2, \dots, \e'_n} \epp
\end{equation}
Then, since the operator $\bR_{1;2, \dots, n}$ is symmetric with respect
to the spin reversal transformation,
\begin{equation}
     \bR_{1;2, \dots, n} = \bJ_{[2,n]} \bR_{1;2, \dots, n} \bJ_{[2,n]}
                           \epc
\end{equation}
the operator $\mathbf{H}_1$ defined by (\ref{H1explicit}) is spin
reversal anti-symmetric
\begin{equation}
     \mathbf{H}_1 = - \bJ \mathbf{H}_1 \bJ \epp
\label{spinrevH1}
\end{equation}
The same is true for the other residues $\mathbf{H}_j$ with $j\ge 2$.
Hence, one concludes that
\begin{equation} \label{spinrevOm2}
     \Om_2 = -\bJ \Omega_2 \bJ \epp
\end{equation}
Moreover, due to the fact that the function $\varphi$ given by eq.\
(\ref{phi}) is an odd function of the magnetic field we have
\begin{equation} \label{hOm2}
     \Om_2 = - \Omega_2 \bigr|_{h \leftrightarrow -h} \epp
\end{equation}
The splitting of the whole operator $\Om$ in equation (\ref{Omega12}) 
into two terms $\Om_1$ and $\Om_2$ seems rather natural because the two
terms are even and odd with respect to the reversal of the spin and the
magnetic field, respectively. 

\section{Examples}
\label{sec:examples}
In this section we present explicit formulae for the density
matrices for $n=1$, $2$ and for some particular matrix elements and
correlation functions for $n=3$. Since the definition of the
operators $\bb$, $\bc$ and $\mathbf{H}$ involves the multiplication
of $2n$ two-by-two matrices and subsequently the calculation of the
traces over $W^+$ or $W^-$, it is already cumbersome to work out by
hand the case $n=2$. We preferred to use a little computer algebra
programme for this task.
\subsection*{The case n = 1}
This case is rather simple because $\Om_1 =0 $ and $\Om = \Om_2$. Since
$\Om^2 = \Om_2^2 = 0$ one should expand the exponent in eq.\
(\ref{main2}) only up to the first order with respect to $\Om$.
A direct calculation shows that the operator $\mathbf{H}_1$ acts on 
the basis elements as follows,
\begin{equation} \label{H1n1}
     \mathbf{H}_1 \bigl({e_1}_{\pm}^{\pm}\bigr) = \pm \frac12 I_1
        \epc \qd
        \mathbf{H}_1 \bigl({e_1}_{\pm}^{\mp}\bigr) = 0 \epp
\end{equation}
Then from (\ref{Omega2Hj}) one obtains $\Om_2$ by multiplying
the above result by $-\ph(\la_1;0)$. It is left to substitute it into 
the formula (\ref{main2}) and take the trace $\frac12\text{tr}_1$.
Finally one obtains the inhomogeneous density matrix
\begin{equation} \label{D1}
     D_1(\la_1|T,h;0) = \frac12\; I_1 - \frac{\varphi(\la_1;0)}2 \s^z_1
                        \epp
\end{equation}
In particular, setting $\la_1=0$ one obtains (see (\ref{oexp}),
(\ref{densinhTh})) for (twice) the magnetization 
\begin{equation}
     \<\s_1^z\>_{T,h} = \tr_1 \bigl( D_1 (T,h) \s_1^z \bigr) =
        - \ph (0;0) \epp
\end{equation}
This result is in full agreement with equation (74) of \cite{GKS05}.

\subsection*{The case n = 2}
This case is already less trivial. First let us calculate $\Om_1$. Using
l'H\^opital's rule and the fact that the functions $\om(\mu_1,\mu_2;0)$
and $\om'(\mu_1,\mu_2;0)$ (recall the definition (\ref{defomprime}) of
$\om'$!) are even and odd, respectively, with respect to the transposition
of $\m_1$ and $\m_2$  (see eq.\ (\ref{omexchange1})) one obtains
\begin{equation} \label{Omega1n2}
     \Om_1 = - \om(\la_1,\la_2;0) \, \Om_1^+
             - \om'(\la_1,\la_2;0) \, \Om_1^- \epc
\end{equation}
where
\begin{align} \label{Omega1+-}
     & \Om_1^+ = \lim_{\a \rightarrow 0} \bigl(
                 \bb_1(\a - 1)\bc_2(\a) \, (\xi_1/\xi_2)^\a
                +\bb_2(\a - 1)\bc_1(\a) \, (\xi_2/\xi_1)^\a \bigr)
                 \epc \notag \\[1ex]
     & \Om_1^- = \lim_{\a \rightarrow 0} \a \bigl(
                 \bb_1(\a - 1)\bc_2(\a) \, (\xi_1/\xi_2)^\a
                -\bb_2(\a - 1)\bc_1(\a) \, (\xi_2/\xi_1)^\a \bigr) \epc
\end{align}
and
\begin{equation}
     \bb_j(\a) = \res_{\z \rightarrow \x_j}
                 \biggl(\bb(\z;\a)\frac{d\z}{\z}\biggr) \epc \qd
     \bc_j(\a) = \res_{\z \rightarrow \x_j}
                 \biggl(\bc(\z;\a)\frac{d\z}{\z}\biggr) \epp
\end{equation}
The result of applying the operators $\Om_1^{\pm}$ to the basis of the
$S^z = 0$ sector is
\begin{align} \label{Omega1+-res}
     & \Om_1^+ \bigl( {e_1}^\e_\e {e_2}^\s_\s \bigr) =
         -\frac{\e\s}4 \cth(\h) \; I_1 I_2 \epc \qd
       \Om_1^+ \bigl( {e_1}^{\mspace{14.mu} \e}_{-\e}
                      {e_2}^{-\e}_{\mspace{14.mu} \e} \bigr) =
       \frac{1}{4} \frac{\ch{(\la_1-\la_2)}}{\sh(\h)} \; I_1 I_2
       \epc \notag \\[1ex]
     & \Om_1^- \bigl( {e_1}^\e_\e {e_2}^\s_\s \bigr) =
       - \frac{\e\s}{4 \h} \cth{(\la_1-\la_2)} \; I_1 I_2 \epc \qd
       \Om_1^- \bigl( {e_1}^{\mspace{14.mu} \e}_{-\e}
                      {e_2}^{-\e}_{\mspace{14.mu} \e} \bigr) =
       \frac{1}{4\h} \frac{\ch(\h)}{\sh{(\la_1-\la_2)}} \; I_1 I_2 \epp
\end{align}
It is clear that
\begin{equation}
     (\Om_1^{\pm})^2 = \Om_1^+ \Om_1^- = \Om_1^- \Om_1^+ = 0
\end{equation}
which implies
\begin{equation} \label{Omega1^2}
     \Om_1^2 = 0 \epp
\end{equation}
Also the symmetry with respect to spin reversal is obvious in the above
explicit formulae (\ref{Omega1+-res}).

Let us proceed with the anti-symmetric part. To obtain $\mathbf{H}_j$ for
$j = 1, 2$ one can either take the corresponding residues in the formula
(\ref{Ha}) or one can use the formulae (\ref{H1explicit}) for $j=1$ and
(\ref{H2explicit}) for $j=2$. The result is
\begin{align} \label{H1res}
     & \mathbf{H}_1 \bigl( {e_1}_\e^\e {e_2}_\s^\s \bigr) =
       \frac{\e}2 \bigl( f_1^{\e\s} (\x_1, \x_2) {e_1}_\e^\e {e_2}_\e^\e +
       f_1^{-\e\s} (\x_1, \x_2) {e_1}_{-\e}^{-\e} {e_2}_{-\e}^{-\e} +
       \notag \\
     & \qd
       + f_2^{\e\s} (\x_1, \x_2) {e_1}_{\e}^{\e} {e_2}_{-\e}^{-\e}
       + f_2^{-\e\s} (\x_1, \x_2) {e_1}_{-\e}^{-\e} {e_2}_{\e}^{\e}
       - \s g_1(\x_1, \x_2) ({e_1}_{+}^{-} {e_2}_{-}^{+}
                              - {e_1}_{-}^{+}{e_2}_{+}^{-})\bigr) \epc 
       \notag \\[1ex]
     & \mathbf{H}_1 \bigl( {e_1}_{\mspace{14.mu} \e}^{-\e}
                           {e_2}_{-\e}^{\mspace{14.mu} \e} \bigr) =
       \frac12 \bigl( q^{-1} f_3^{\e} (\x_1, \x_2)
                        {e_1}_\e^\e {e_2}_{\e}^{\e}
                    + q f_3^{-\e} (\x_1, \x_2) {e_1}_{-\e}^{-\e}
                                               {e_2}_{-\e}^{-\e} +
       \notag \\
     & \qd
       + \e q^{-1} g_2^+ (\x_1, \x_2) {e_1}_\e^\e {e_2}_{-\e}^{-\e}
       - \e q g_2^- (\x_1, \x_2) {e_1}_{-\e}^{-\e} {e_2}_\e^\e
       + g_3 (\x_1, \x_2) ({e_1}_+^- {e_2}_-^+
       - {e_1}_-^+ {e_2}_+^-) \bigr)
\end{align}
and
\begin{align} \label{H2res}
     & \mathbf{H}_2 \bigl( {e_1}_\e^\e {e_2}_\s^\s \bigr) =
       \frac{\s}2 \bigl( f_1^- (\x_1, \x_2) {e_1}_\e^\e \, I_2 +
        f_1^+ (\x_1, \x_2) {e_1}_{-\e}^{-\e} \, I_2 \bigr) \epc \notag \\
     & \mathbf{H}_2 \bigl( {e_1}_{\mspace{14.mu} \e}^{-\e}
                           {e_2}_{-\e}^{\mspace{14.mu} \e}\bigr) =
       - \frac{\e}2 \bigl( q^{-1} f_3^+ (\x_1, \x_2) {e_1}_\e^\e \, I_2 +
         q f_3^- (\x_1, \x_2) {e_1}_{-\e}^{-\e} \, I_2 \bigr) \epc
\end{align}
where
\begin{align} \label{f_i}
     & f_1^+ (\x_1, \x_2) := \frac1{1 - \x_1^2/\x_2^2} \epc \notag \\
     & f_2^+ (\x_1, \x_2) := \frac{(q-q^{-1})^2 + (1 - \x_1^2/\x_2^2)^2}
                                  {(1 - \x_1^2/\x_2^2)
                                   (q \x_1/\x_2 - q^{-1} \x_2/\x_1)
                                   (q \x_2/\x_1 - q^{-1} \x_1/\x_2)}
       \epc \notag \\
     & f_3^+ (\x_1, \xi_2) := \frac1{\x_1/\x_2 - \x_2/\x_1}
       \epc \notag \\
     \text{and} \qd & f_i^- (\x_1, \x_2) := f_i^+ (\x_2, \x_1) \epp
\end{align}
\begin{align} \label{g_i}
     & g_1 (\x_1, \x_2) := \frac{(\x_1/\x_2 + \x_2/\x_1)(q-q^{-1})}
                                {(q \x_1/\x_2 - q^{-1} \x_2/\x_1)
                                 (q \x_2/\x_1 - q^{-1} \x_1/\x_2)} \epc
                           \notag \\
     & g_2^{\pm} (\x_1, \x_2) := \frac{(q-q^{-1})^2 + q^{\pm 2}
                                       (\x_1/\x_2 - \x_2/\x_1)^2}
                                      {(\x_1/\x_2 - \x_2/\x_1)
                                       (q \x_1/\x_2 - q^{-1} \x_2/\x_1)
                                       (q \x_2/\x_1 - q^{-1} \x_1/\x_2)}
                                 \epc \notag \\
     & g_3 (\x_1, \x_2) := \frac{q^2 - q^{-2}}
                                {(q \x_1/\x_2 - q^{-1} \x_2/\x_1)
                                 (q \x_2/\x_1 - q^{-1} \x_1/\x_2)} \epp
\end{align}
The anti-symmetry of the operators $\mathbf{H}_1$ and $\mathbf{H}_2$
with respect to the spin reversal transformation is evident in the
above formulae.

Also one can directly verify that 
\begin{equation} \label{algH1H2}
     \mathbf{H}_1^2 = \mathbf{H}_2^2 = \mathbf{H}_1 \mathbf{H}_2
                                     + \mathbf{H}_2 \mathbf{H}_1 = 0
\end{equation}
and
\begin{equation} \label{HjOmega1}
     \mathbf{H}_j \Om_1 + \Om_1 \mathbf{H}_j = 0 \epc \qd j = 1, 2 \epp
\end{equation}
This means that the operator $\Om_2$ which is
\begin{equation} \label{Omega2n2}
     \Om_2 = - \ph (\la_1;0) \mathbf{H}_1 - \ph(\la_2;0) \mathbf{H}_2
\end{equation}
satisfies
\begin{equation} \label{Omega1Omega2}
     \Omega_2^2 = \Om_1 \Om_2 + \Om_2 \Om_1 = 0
\end{equation}
From this follows that
\begin{equation} \label{Omegan2nilpot}
     \Om^2 = 0
\end{equation}
and the expansion of the exponent in the formula (\ref{main2}) extends
only up to the first order in powers of $\Om$.

Therefore in order to compute the elements of the density matrix we
need to calculate the traces
\begin{equation} \label{D2}
     {D_2\,}^{\e_1,\e_2}_{\s_1,\s_2} (\la_1,\la_2|T,h;0) =
        \frac14 \tr_1 \tr_2 \bigl[ (\id + \Om_1 + \Om_2)
        \bigl({e_1}_{\e_1}^{\s_1} {e_2}_{\e_2}^{\s_2} \bigr)\bigr] \epp
\end{equation}
For this purpose we have to use the formulae (\ref{Omega1n2}),
(\ref{Omega1+-res})  and (\ref{Omega2n2}), (\ref{H1res}), (\ref{H2res}).
The result decomposes as follows,
\begin{equation} \label{D2res}
     D_2 (\la_1,\la_2|T,h;0) = D_2^{\text{even}} (\la_1,\la_2) +
                               D_2^{\text{odd}} (\la_1,\la_2) \epc
\end{equation}
where $D_2^{\text{even}}$ and $D_2^{\text{odd}}$ are $4\times 4$ matrices,
\begin{multline} \label{D2even}
     D_2^{\text{even}} (\la_1,\la_2) =
        \frac14 \; I\otimes I
	+ \4 \biggl[ \cth(\h) \om(\la_1,\la_2;0) +
                     \frac{\cth(\la_1-\la_2)}{\h} \om'(\la_1,\la_2;0)
		     \biggr] \s^z \otimes \s^z \\[1ex]
	- \4 \biggl[ \frac{\ch(\la_1 - \la_2)}{\sh(\h)}
	             \om(\la_1,\la_2;0) +
		     \frac{\ch(\h)}{\h \sh(\la_1 - \la_2)}
		     \om'(\la_1,\la_2;0) \biggr]
                     (\s^+ \otimes \s^- + \s^- \otimes \s^+)
\end{multline}
and
\begin{multline} \label{D2odd}
     D_2^{\text{odd}} (\la_1,\la_2) =
        - \frac{\ph(\la_1;0)}4 \s^z\ \otimes I
        - \frac{\ph(\la_2;0)}4 I \otimes \s^z \\[1ex]
        - \frac{\sh(\h) \bigl(\ph(\la_1;0) - \ph(\la_2;0)\bigr)}
	       {4 \sh(\la_1-\la_2)}
          (\s^+ \otimes \s^- - \s^- \otimes \s^+) \epp
\end{multline}
The homogeneous limit $\la_1, \la_2 \rightarrow 0$ can be readily taken.
We obtain the density matrix for $n = 2$,
\begin{multline} \label{densmat2}
     D_2 (T,h) = \4 \biggl[
        I \otimes I - \ph (\s^z \otimes I + I \otimes \s^z)
        - \sh(\h) \ph_x (\s^+ \otimes \s^- - \s^- \otimes \s^+) \\[1ex]
	+ \left(\cth (\h) \om + \frac{\om_x'}\h\right)
                            \s^z \otimes \s^z
        - \left(\frac{\om}{\sh(\h)} + \frac{\ch(\h) \om_x'}{\h}\right)
               (\s^+ \otimes \s^- + \s^- \otimes \s^+)
	\biggr] \epc
\end{multline}
where we introduced the shorthand notation
\begin{equation}
     \ph = \ph(0;0) \epc \qd
     \ph_x = \6_\la \ph(\la;0) \Bigr|_{\la = 0} \epc \qd
     \om = \om(0,0;0) \epc \qd
     \om_x' = \6_{\la_1} \om'(\la_1, \la_2; 0) \Bigr|_{\la_1, \la_2 = 0}
              \epp
\end{equation}
The density matrix (\ref{densmat2}) can now be used to obtain any
two-site correlation function, e.g.,
\begin{subequations}
\begin{align}
     \<\s_1^z \s_2^z\>_{T,h} & =
        \tr_{12} \bigl( D_2 (T,h) \s_1^z \s_2^z \bigr) =
        \cth (\h) \om + \frac{\om_x'}\h \epc \\
     \<\s_1^x \s_2^x\>_{T,h} & =
        \tr_{12} \bigl( D_2 (T,h) \s_1^x \s_2^x \bigr) =
        - \frac{\om}{2 \sh(\h)} - \frac{\ch(\h) \om_x'}{2 \h} \epp
\end{align}
\end{subequations}
\subsection*{The case n = 3}
The explicit forms of $\Omega_j$ or $\mathbf{H}_j$ are already quite
involved for $n=3$. We shall not present the exhausting list of
matrix elements, but rather restrict ourselves to some examples of
physical interest.

We introduce shorthand notations
\begin{align}
     \mathfrak{d}^{\ve_1,\ve_2}_1 &
        = f^{\ve_1}_1 (\xi_2,\xi_3) f^{\ve_2}_1 (\xi_3,\xi_1) \epc
      & \mathfrak{d}^{\ve_1,\ve_2}_2
      & = f^{\ve_1}_1 (\xi_1,\xi_2) f^{\ve_2}_1(\xi_3,\xi_1) \epc
          \notag \\
        \mathfrak{d}^{\ve_1,\ve_2}_3
      & = f^{\ve_1}_1 (\xi_1,\xi_2) f^{\ve_2}_1 (\xi_2,\xi_3) \epc
          \notag \\
        \mathfrak{t}^{\ve_1,\ve_2,\ve_3}
      & = f^{\ve_1}_1 (\xi_1,\xi_2) f^{\ve_2}_1 (\xi_2,\xi_3)
          f^{\ve_3}_1 (\xi_3,\xi_1) \epp
\end{align}
Using these symbols, the longitudinal correlation is represented
rather compactly. In the inhomogeneous case we find
\begin{align}
     \tr_{123} \bigl( D_3 (\la_1, & \la_2, \la_3|T,h;0)
                      \s_1^z \s_3^z \bigr) \notag \\[1ex]
     = & \tgh(\h) \bigl(\mathfrak{d}^{++}_1 + \mathfrak{d}^{--}_1 \bigr)
	              \omega(\lambda_1,\lambda_2;0)
         + \tgh(\h) \bigl(\mathfrak{d}^{++}_2
                          + \mathfrak{d}^{--}_2 \bigr)
	                    \omega(\lambda_2,\lambda_3;0) \notag \\[1ex]
       & + \bigl( 2 \cth(2\h) (\mathfrak{d}^{++}_3
               + \mathfrak{d}^{--}_3) + \cth(\h) (\mathfrak{d}^{+-}_3
	       + \mathfrak{d}^{-+}_3) \bigr)
	         \omega(\lambda_3,\lambda_1;0) \notag \\
       & - \frac{4 \sh^2 (\h)}{\eta} \mathfrak{t}^{+++}
           \bigl( \omega^{\prime} (\lambda_1,\lambda_2;0)
	        + \omega^{\prime} (\lambda_2,\lambda_3;0) \bigr)
		  \notag \\ 
       & - \frac{1}{\eta} \bigl( 4 \ch^2(\h) \mathfrak{t}^{+++}
         - (\mathfrak{t}^{++-}+\mathfrak{t}^{-+-}+\mathfrak{t}^{+--})
	   \notag \\
       & \phantom{vvvvvvvvvvvvvvvvvvvvv}
         + \mathfrak{t}^{+-+}+\mathfrak{t}^{-++}+\mathfrak{t}^{--+}
	    \bigr) \omega^{\prime} (\lambda_3,\lambda_1;0) \epp
\end{align}
Taking the homogeneous limit we arrive at
\begin{equation}
     \<\s^z_1 \s^z_3\>_{T, h} =
        2 \cth(2\h) \omega + \4 \tgh(\h) \omega_{xx}
	- \2 \tgh(\h) \omega_{xy}
        + \frac{\omega^{\prime}_x}{\eta}
        - \frac{\sh^2(\h)}{4 \eta} \omega^{\prime}_{xxy} \epp
\end{equation}
By $x$ and $y$ we denote the derivatives with respect to first and
second argument taken at zero. The same limit for the transverse
correlation reads as follows.
\begin{multline}
     \<\s^+_1 \s^-_3  + \s^-_1 \s^+_3 \>_{T,h} \\ =
        - \frac{1}{\sh(2 \h)} \omega
	+ \frac{\ch(2\h) \tgh(\h)}{4} \omega_{xy}
	- \frac{\ch(2\h) \tgh(\h)}{8} \omega_{xx}
        - \frac{\ch(2\h)}{2\h} \omega^{\prime}_x
	+ \frac{\sh^2(\h)}{8\h} \omega^{\prime}_{xxy} \epp
\end{multline}
The rational limit in the last two equations is not easy. Using the
high-temperature expansion we checked to ${\cal O} (1/T)$ that it
coincides with our previous result \cite{BGKS06} for the XXX chain.

As a last example we show the emptiness formation probability in
the inhomogeneous case,
\begin{multline}
     D^{+++}_{+++} (\la_1, \la_2, \la_3|T,h;0)
        = \frac{1}{8} + \frac{1}{8}
	  \bigl( - \varphi(\lambda_3;0)
	         + C_1(\xi_1,\xi_2,\xi_3) \omega(\lambda_1,\lambda_2;0)\\
		 + C_2 (\xi_1,\xi_2,\xi_3)
		        \omega^{\prime}(\lambda_1,\lambda_2;0)
        + C_3(\xi_1,\xi_2,\xi_3) \omega(\lambda_1,\lambda_2;0)
	    \varphi(\lambda_3;0) \\[1ex]
	  + C_4(\xi_1,\xi_2,\xi_3)
	    \omega^{\prime}(\lambda_1,\lambda_2;0) \varphi(\lambda_3;0)
        + {\text {cyclic permutations } } \bigr) \epp
\end{multline}
Here the coefficients are given as follows,
\begin{align}
     C_1(\xi_1,\xi_2,\xi_3) & =
        \bigl( 2 \cth(2\h)
	     ( \mathfrak{d}^{++}_1 + \mathfrak{d}^{--}_1 )
	 +   \cth(\h)
	     ( \mathfrak{d}^{+-}_1 + \mathfrak{d}^{-+}_1 ) \bigr) \epc
	     \notag \\
     C_2 (\xi_1,\xi_2,\xi_3) & =
        \frac{1}{\eta} \bigl(
	- 2 \ch(2\h) \, \mathfrak{t}^{+++}
	- ( \mathfrak{t}^{++-}+ \mathfrak{t}^{+-+}+ \mathfrak{t}^{+--})
        + \mathfrak{t}^{--+}+ \mathfrak{t}^{-+-}+\mathfrak{t}^{-++}
          \bigr) \epc \notag \\
     C_3 (\xi_1,\xi_2,\xi_3) & =
        \cth(\h) \bigl( 
        2 (\mathfrak{d}^{++}_1 + \mathfrak{d}^{--}_1)
	- (\mathfrak{d}^{+-}_1+\mathfrak{d}^{-+}_1) \bigr) \epc \notag \\
     C_4 (\xi_1,\xi_2,\xi_3) & =
        \frac{1}{\eta} \bigl(
        - 4 \ch^2 (\h) \, \mathfrak{t}^{+++}
	+ \mathfrak{t}^{++-}+\mathfrak{t}^{+-+}+\mathfrak{t}^{-++}
        - (\mathfrak{t}^{--+}+\mathfrak{t}^{-+-}+\mathfrak{t}^{+--})
	\bigr).
\end{align}
The homogeneous limit is left as an exercise to the reader.
\section{Conclusions}
\label{sec:concl}
In an attempt to generalize the recent results \cite{BJMST06b,BJMST07app}
on the factorization of the ground state correlation functions of the
XXZ chain to include finite temperatures and a finite longitudinal
magnetic field we have constructed a conjectural exponential formula
(\ref{Omega12}), (\ref{Omega1and2}) for the density matrix. The main
steps in our work were the construction of the operator $\mathbf{H}$,
eq.\ (\ref{H}), which takes care of the modification of the algebraic
part of the exponential formula in the presence of a magnetic field, and
of the functions $\ph$ and $\om$, eqs.\ (\ref{phi}), (\ref{om}),
which allowed us to give a description of the physical part in close
analogy to \cite{BJMST06b,BJMST07app}. In the limit $T, h \rightarrow 0$
our conjecture reduces to the result of \cite{BJMST06b,BJMST07app},
even for finite $\a$. It also trivializes in the expected way as
$T \rightarrow \infty$. We tested our conjecture against the multiple
integral formula (\ref{densint}) by direct comparison for $n = 2$
(see appendix \ref{app:doubleint}) and by comparison of the high
temperature expansion data for $n = 3$ and $n = 4$. Judging from
our experience with the isotropic case \cite{DGHK07} we find it likely
that very similar formulae also hold in the finite length case and that
the only modifications necessary to cover this case are a restriction
of $\Om$ to the finite length $L$ of the chain and a change of the
auxiliary function from (\ref{nlie}) to (\ref{nliefini}).
\\[1ex]{\bf Acknowledgement.}
The authors are grateful to M. Jimbo, T. Miwa, F. Smirnov and to
Y. Takeyama for stimulating discussions. HB was supported by the
RFFI grant \#04-01-00352. FG is grateful to Shizuoka University for
hospitality. JS and HB acknowledge partial financial support
by the DFG-funded research training group 1052 -- `Representation
Theory and its Applications', JS also by a Grand-in-Aid for Scientific
Research \#17540354 from the Ministry of Education of Japan.

\clearpage

{\appendix
\Appendix{Proof of equations (\ref{Qexplicitinhom}) and
(\ref{Q1explicit})}
\label{app:proofs}
\noindent
Here we outline the proof of eqs.\ (\ref{Qexplicitinhom}) and
(\ref{Q1explicit}). Our starting point is eq.\ (\ref{Q+}). Since we
work in the sector $S^z = 0$, we have to set ${\mathbb S} = 0$. Then
\begin{multline} \label{Q+basis}
     \bigl[ \bQ^+ (\z; \a) \bigl]^{\s_1,\cdots,\s_n; \e_1,\cdots,\e_n}_%
                  {\s'_1,\cdots,\s'_n;\;\e'_1,\cdots,\e'_n} = 
        (1-q^{2\al}) \\[1ex] \times
        \tr^+_A \( \bigl( {L^+_A(\z/\x_1)^{-1} \bigr)}_{\s'_1}^{\s_1}
        \dots \bigl( {L^+_A(\z/\x_n)^{-1} \bigr)}_{\s'_n}^{\s_n}
        \bigl( {L^+_A(\z/\xi_n) \bigr)}_{\e_n}^{\e'_n}
        \dots \bigl( {L^+_A(\z/\xi_1) \bigr)}_{\e_1}^{\e'_1}
        q^{2\al D_A} \) \epc
\end{multline}
where $\bigl({L^+_A(\z)\bigr)}_{\e}^{\e'}$ are the matrix elements of 
the $L$-operator (\ref{L+}),
\begin{equation} \label{elemL+}
     \bigl({L^+_A(\z)\bigr)}_{\e}^{\e'} =
        \i \z ^{-\frac 1 2}q^{-\frac 1 4}
        \begin{pmatrix}
           q^{D_A} & -\z a_A^* q^{-D_A}\\[5pt]
           - \z  a_A q^{D_A} & q^{-D_A} - \z ^2q^{D_A+2}
        \end{pmatrix}_{\e,\e'}
\end{equation}
and
\begin{equation} \label{elemL+inv}
     \bigl({{L^+_A(\z)}^{-1}\bigr)}_{\e}^{\e'} =
        \frac{\i \z ^{-\frac 1 2} q^{\frac 1 4}}{\z - \z^{-1}}
        \begin{pmatrix}
           q^{-D_A} - \z^2 q^{D_A} & \z q^{-D_A} a_A^*\\[5pt]
           \z q^{D_A} a_A & q^{D_A}
        \end{pmatrix}_{\e,\e'} \epp
\end{equation}
The main observation is that for the computation of the limit $\a
\rightarrow 0$ of eq.\ (\ref{Q+basis}) it is enough to substitute there
$L^+_A$ and $(L^+_A)^{-1}$ by $\tilde L^+_A$ and $(\tilde L^+_A)^{-1}$
with\footnote{Strictly speaking the operators $\tilde L^+_A$ and
$(\tilde L^+_A)^{-1}$ are not inverse to each other any longer.}
\begin{equation}
     \bigl( {\tilde L^+_A(\z) \bigr)}_{\e}^{\e'} =
        \i \z ^{-\frac 1 2} q^{-\frac 1 4}
        \begin{pmatrix}
           q^{D_A} & - \z a_A^* q^{-D_A}\\[5pt]
           - \z  a_A q^{D_A} & q^{-D_A}
        \end{pmatrix}_{\e,\e'} =
        \i \z^{-\frac 1 2} q^{-\frac 1 4} \; \e \e' \;
        \z^{\frac{1 - \e \e'}2} a_A^{-\frac{\e - \e'}2} q^{\e' D_A} 
\end{equation}
and
\begin{equation}
     \bigl( {\tilde L^+_A(\z)^{-1} \bigr)}_{\e}^{\e'} =
        \frac{\i \z^{-\frac 1 2}q^{\frac 1 4}}{\z - \z^{-1}}
        \begin{pmatrix}
           q^{-D_A} & \z q^{-D_A} a_A^*\\[5pt]
           \z q^{D_A} a_A & q^{D_A}
        \end{pmatrix}_{\e,\e'} =
        \frac{\i \z^{-\frac 1 2} q^{\frac 1 4}}{\z - \z^{-1}}\;
        \z^{\frac{1 - \e \e'}2} q^{-\e D_A} a_A^{-\frac{\e - \e'}2} \epc
\end{equation}
where we set
\begin{equation}
     a_A^+ = a_A \epc \qd a_A^- = a_A^\ast \epp
\end{equation}
In this notation the algebra (\ref{alg}) looks very simple
\begin{equation} \label{permute}
     a_A^{\e} q^{\e' D_A} = q^{\e\e'} q^{\e' D_A} a_A^{\e} \epp
\end{equation}
The reason is as follows.
Let us first formally substitute $\tilde L^+_A$ and $(\tilde L^+_A)^{-1}$
for $L^+_A$ and $(L^+_A)^{-1}$ into the right hand side of eq.\
(\ref{Q+basis}),
\begin{multline} \label{expwithtilde}
     (1-q^{2\al}) \prod_{j=1}^n
        \frac{\e_j \e'_j\; (\z/\x_j)^{-\frac12(\e_j\e'_j + \s_j\s'_j)}}
             {\z/\x_j - \x_j/\z}
        \tr^+_A \bigl(q^{-\s'_1 D_A} a_A^{-\frac12(\s'_1 - \s_1)}
        \dots q^{-\s'_n D_A} a_A^{-\frac12 (\s'_n - \s_n)} \\ \times
     a_A^{-\frac12(\e_n - \e'_n)} q^{\e'_n D_A} \dots
     a_A^{-\frac12(\e_1 - \e'_1)} q^{\e'_1 D_A} q^{2\al D_A} \bigr) \epp
\end{multline}
We do not write the $\delta$'s like in the right hand side of
(\ref{Qexplicitinhom}) which reflect the fact that we are in the spin-0
sector. Let us just imply that they are there.

Let us formally ignore all $a_A^{\pm}$ inside the trace  here. Then the
total degree of $q^{D_A}$ is zero because $\sum_j{\e'_j}=\sum_j{\s'_j}$,
and only $q^{2\al D_A}$ is left, which produces a term $1/(1-q^{2\al})$
after taking the trace over the oscillator space $A$. Since the
differences between $L^+_A$ and $\tilde L^+_A$ and between
$(L^+_A)^{-1}$ and $(\tilde L^+_A)^{-1}$ contain only positive powers of
$q^{D_A}$, the insertion of such terms does not change that most singular
term $1/(1-q^{2\al})$ when $\al\rightarrow 0$. Therefore we can ignore
those differences when calculating the limit  $\al\rightarrow 0$. 

One more observation is about the contribution coming from the terms
containing $a_A^{\pm}$. Suppose we had just
\[
     \tr^+_A \( a_A^{\e_1} \dots a_A^{\e_{2n}} q^{2\al D_A} \)
\] 
with $\e_1 + \dots + \e_{2n} = 0$. Then, using the algebra (\ref{alg}) 
we would conclude that again the most singular term would be
$1/(1-q^{2\al})$ as a result of taking the trace. It means that if one
succeeds in collecting all $a_A^{\pm}$ then one can replace them by
1 without any change in the most singular term. The first conclusion
obtained from the above is that the limit $\a \rightarrow 0$ of the
expression (\ref{expwithtilde}) gives us the limit $\al\rightarrow 0$
of eq.\ (\ref{Q+basis}). Second, in order to calculate it we have to
collect all $a_A^{\pm}$ inside the trace (\ref{expwithtilde}) using
the algebra (\ref{permute}) in one place, say in the place of the symbol
$\times$ in (\ref{expwithtilde}). If we do this and afterwards ignore
the product of all $a_A^{\pm}$ following the above arguments, then we
can easily take the limit $\al\rightarrow 0$ and come to the formula
(\ref{Qexplicitinhom}). Similar arguments may be applied when treating
the $\a \rightarrow 0$ limit of the formula (\ref{Q-}).

Now we outline the derivation of the formula (\ref{Q1explicit}). 
When calculating the residue at $\x_1$ which is implied in eq.\
(\ref{H1explicit}) one obtains
\begin{multline} \label{Q1basis}
     \bigl[ \bQ_{\s_1}^{\e_1} \bigl]^%
            {\s_2, \dots, \s_n; \e_2,\dots,\e_n}_%
            {\s'_2, \dots, \s'_n; \e'_2, \dots, \e'_n} =
     \lim_{\a \rightarrow 0} \res_{\z = \x_1} \tr^+_A
     \( \bigl( L^+_A(\z/\x_1)^{-1} \bigr)_{\s}^{\s_1}
        \bigl( L^+_A(\z/\x_2)^{-1} \bigr)_{\s'_2}^{\s_2} \dots
        \bigl( L^+_A(\z/\x_n)^{-1} \bigr)_{\s'_n}^{\s_n} \right. \\
     \left.
     \bigl( L^+_A(\z/\xi_n)\bigr)_{\e_n}^{\e'_n} \dots
     \bigl( L^+_A(\z/\xi_2)\bigr)_{\e_2}^{\e'_2}
     \bigl( L^+_A(\z/\xi_1)\bigr)_{\e_1}^{\s} q^{2\al D_A} \) \epc
\end{multline}
where summation over $\s$ is implied. The pole at $\z = \x_1$ originates
from the $L$-operators with argument $\z/\xi_1$. We use the cyclicity
of the trace and directly verify that
\begin{multline} \label{LL}
     \res_{\z = \x_1} \bigl( L^+_A(\z/\x_1)\bigr)_{\e_1}^{\s}
        q^{2\a D_A} \bigl( L^+_A(\z/\xi_1)^{-1} \bigr)_{\s}^{\s_1} \\ =
     - \frac{q^{\al}-q^{-\al}}2
     \biggl[ \begin{pmatrix} 1 \\ -a_A \end{pmatrix}
             a_A^{\ast}
             \begin{pmatrix} q^{-\al} a_A, q^{\al} \end{pmatrix}
     \biggr]_{\e_1,\s_1}q^{2\al D_A} \epp
\end{multline}
Implying that we need to calculate the limit $\al\rightarrow 0$ in the
end we may set
\begin{equation} \label{LL1}
     (1 - q^{\a}) \biggl[ \begin{pmatrix} 1\\ - a_A \end{pmatrix} 
        \begin{pmatrix} 1, a_A^{\ast} \end{pmatrix} \biggr]_{\e_1,\s_1}
        q^{2\al D_A} = (1-q^{\al}) \e_1 a^{-\frac12(\e_1 - \s_1)}
                        q^{2\al D_A}
\end{equation}
on the right hand side of (\ref{LL}). Thus, we come to the conclusion
that the right hand side of (\ref{Q1basis}) is equal to
\begin{multline} \label{Q1basisa}
     \lim_{\a \rightarrow 0} \left[
        (1 - q^\a) \tr^+_A \bigl(
        \bigl( L^+_A(\x_1/\x_2)^{-1} \bigr)_{\s'_2}^{\s_2} \dots
        \bigl( L^+_A(\x_1/\x_n)^{-1} \bigr)_{\s'_n}^{\s_n} \right. \\
     \left.
     \times \bigl( L^+_A(\x_1/\x_n) \bigr)_{\e_n}^{\e'_n} \dots
        \bigl( L^+_A(\x_1/\x_2) \bigr)_{\e_2}^{\e'_2}
        \e_1 a^{-\frac12(\e_1 - \s_1)} q^{2\a D_A} \bigr) \right] \epp
\end{multline}
Finally we can apply to eq.\ (\ref{Q1basisa}) the same trick as
described above in order to get the formula~(\ref{Q1explicit}).

\Appendix{Factorization of the double integral}
\label{app:doubleint}
\noindent
In this appendix we show that our conjectured formula for the density
matrix for $n = 2$, eqs.\ (\ref{D2res})-(\ref{D2odd}), coincides with
the double integral, eq.\ (\ref{densint}) for $n = 2$. The density
matrix for $n = 2$ has six non-vanishing elements. In this appendix
we will denote it by $D$ rather than by $D_2$ and suppress the
temperature, magnetic field and $\a$ dependence of the matrix elements for
short. Using the Yang-Baxter algebra and reduction we find four
independent relations between the six non-vanishing matrix elements
of $D$,
\begin{align} \label{densmatrel2}
     & D^{+-}_{+-} (\la_1, \la_2)
        = D^+_+ (\la_1) - D^{++}_{++} (\la_1, \la_2) \epc \qd
       D^{-+}_{-+} (\la_1, \la_2)
        = D^+_+ (\la_2) - D^{++}_{++} (\la_1, \la_2) \epc \notag \\[1ex]
     & D^{--}_{--} (\la_1, \la_2) = D^{++}_{++} (\la_1, \la_2)
        - D^+_+ (\la_1) - D^+_+ (\la_2) + 1 \epc \notag \\[1ex]
     & D_{+-}^{-+} (\la_1, \la_2) - D_{-+}^{+-} (\la_1, \la_2) =
       \frac{\sh(\h) \bigl(D^+_+ (\la_1) - D^+_+ (\la_2)\bigr)}
            {\sh(\la_1 - \la_2)} \epp
\end{align}
Inserting these relations into $D (\la_1, \la_2| T, h; 0)$ we obtain
\begin{multline} \label{d2expand}
     D (\la_1, \la_2| T, h; 0) =
        \4 I \otimes I + \4 \bigl(2 D^+_+ (\la_1) - 1\bigr) \s^z \otimes I
        + \4 \bigl(2 D^+_+ (\la_2) - 1\bigr) I \otimes \s^z \\
        + \4 \bigl(4 D^{++}_{++} (\la_1, \la_2)
              - 2 D^+_+ (\la_1) - 2 D^+_+ (\la_2) + 1\bigr)
              \s^z \otimes \s^z  \\
        + \2 \bigl(D^{-+}_{+-} (\la_1, \la_2)
                   + D^{+-}_{-+} (\la_1, \la_2)\bigr)
             (\s^+ \otimes \s^- + \s^- \otimes \s^+) \\
        + \frac{\sh(\h)}{2 \sh(\la_1 - \la_2)}
          \bigl(D^+_+ (\la_1) - D^+_+ (\la_2)\bigr)
          (\s^+ \otimes \s^- - \s^- \otimes \s^+) \epc
\end{multline}
and we are left with the problem of expressing the one-point function
$D^+_+ (\la)$ and the two-point functions $D^{++}_{++} (\la_1, \la_2)$
and $D^{-+}_{+-} (\la_1, \la_2) + D^{+-}_{-+} (\la_1, \la_2)$ in terms
of $\ph$ and $\om$.

Comparing (\ref{densint}) for $n = 1$ with the definition (\ref{phi})
of our function $\ph$ we find the relation
\begin{equation} \label{onepointapp}
     2 D^+_+ (\la) = 1 - \ph(\la; 0)
\end{equation}
for the one-point function.

In order to simplify our task for the two-point functions we introduce
the quantum group invariant combination \cite{BJMST06}
\begin{multline} \label{defdq}
     D_q (\la_1, \la_2) \\ = 
        \re^{\la_1 - \la_2} D^{+-}_{-+} (\la_1, \la_2)
        + \re^{\la_2 - \la_1} D^{-+}_{+-} (\la_1, \la_2)
        - \re^{- \h} D^{+-}_{+-} (\la_1, \la_2)
        - \re^\h D^{-+}_{-+} (\la_1, \la_2) \epp
\end{multline}
Using again (\ref{densmatrel2}) we obtain the relation
\begin{multline} \label{dpmmp}
     \ch(\la_1 - \la_2) \bigl( D^{-+}_{+-} (\la_1, \la_2)
        + D^{+-}_{-+} (\la_1, \la_2) \bigr) \\ = D_q (\la_1, \la_2)
        + \ch(\h) \bigl( D^+_+ (\la_1) + D^+_+ (\la_2)
                         - 2 D^{++}_{++} (\la_1, \la_2) \bigr) \epp
\end{multline}
Hence, in order to determine the density matrix for $n = 2$, it suffices
to calculate $ D_q (\la_1, \la_2)$ and $D^{++}_{++} (\la_1, \la_2)$ 
from the double integrals.

Let us start with the simpler case $D_q (\la_1, \la_2)$. Inserting
(\ref{densint}) into the definition (\ref{defdq}) we find
\begin{equation} \label{dqint}
     D_q (\la_1, \la_2) = 
        \int_{\cal C} \frac{\rd \om_1}{2 \p \i (1 + \fa(\om_1))}
        \int_{\cal C} \frac{\rd \om_2}{2 \p \i (1 + \faq(\om_2))}
        \det \bigl[ - G(\om_j, \la_k; 0) \bigr] r(\om_1, \om_2) \epc
\end{equation}
where
\begin{equation}
     r(\om_1, \om_2) =
        - \frac{\re^{\la_1 + \la_2}
	        \sh(\la_1 - \la_2 + \h) \sh(\la_1 - \la_2 - \h)}
               {\re^{\om_1 + \om_2}
	        \sh(\la_1 - \la_2) \sh(\om_1 - \om_2 - \h)} \epp
\end{equation}
Using the simple relation
\begin{equation}
     \frac{1}{1 + \fa(\om)} + \frac{1}{1 + \faq(\om)} = 1
\end{equation}
we can rewrite (\ref{dqint}) as
\begin{multline}
     D_q (\la_1, \la_2) = 
        \int_{\cal C} \frac{\rd \om}{2 \p \i (1 + \fa(\om))}
        \bigl( r(\om, \la_1) G(\om, \la_2; 0)
	     - r(\om, \la_2) G(\om, \la_1; 0) \bigr) \\[1ex] - \2
        \int_{\cal C} \frac{\rd \om_1}{2 \p \i (1 + \fa(\om_1))}
        \int_{\cal C} \frac{\rd \om_2}{2 \p \i (1 + \fa(\om_2))}
        \det \bigl[ - G(\om_j, \la_k; 0) \bigr]
        \bigl( r(\om_1, \om_2) - r(\om_2, \om_1) \bigl) \epp
\end{multline}
The first term on the right hand side is already a single integral.
For the second term we observe that
\begin{equation}
     - \2 \bigl( r(\om_1, \om_2) - r(\om_2, \om_1) \bigl)
        = \frac{\ch(\h) \bigl( \re^{- 2 \om_1} - \re^{- 2 \om_2} \bigr)
                \sh(\la_1 - \la_2 + \h) \sh(\la_1 - \la_2 - \h)}
               {\bigl( \re^{- 2 \la_1} - \re^{- 2 \la_2} \bigr)
                \sh(\om_1 - \om_2 + \h) \sh(\om_1 - \om_2 - \h)} \epp
\end{equation}
The $\om$-dependent terms in the denominator are proportional to
the kernel in the integral equation (\ref{G}) for $\a = 0$, and the
numerator is a sum of a function of $\om_1$ and a function of $\om_2$.
Hence, the double integral can be reduced to single integrals by
means of the integral equation (\ref{G}). Collecting the resulting
terms and inserting the definition (\ref{psi}) of our function $\ps$
we arrive at
\begin{equation} \label{dqpsi}
     D_q (\la_1, \la_2) = 
        \frac{\sh(\la_1 - \la_2 + \h) \sh(\la_1 - \la_2 - \h)}
             {2 \sh(\h)} \ps(\la_1, \la_2; 0) \epp
\end{equation}

Let us proceed with the calculation of $D^{++}_{++} (\la_1, \la_2)$
which according to (\ref{densint}) is equal to
\begin{multline} \label{pppint}
     D^{++}_{++} (\la_1, \la_2) = \lim_{\a \rightarrow 0}
        \int_{\cal C}
	   \frac{\rd \om_1 \; \re^{-\a \h}}{2 \p \i (1 + \fa(\om_1))}
        \int_{\cal C}
	   \frac{\rd \om_2 \; \re^{-\a \h}}{2 \p \i (1 + \fa(\om_2))} \\
        \det \bigl[ - G(\om_j, \la_k; \a) \bigr]
	\underbrace{\frac{\sh(\om_1 - \la_1 - \h) \sh(\om_2 - \la_2)}
	                 {\sh(\la_2 - \la_1) \sh(\om_1 - \om_2 - \h)}}_%
			 {= s(\om_1, \om_2)} \epp
\end{multline}
Because of the antisymmetry of the determinant we may replace
$s(\om_1, \om_2)$ with
\begin{multline*}
     \2 \bigl( s(\om_1, \om_2) - s(\om_2, \om_1) \bigr) \\ =
        \frac{\ch(\h) \bigl( \sh(2 \om_2 - \la_1 - \la_2 - \h)
                             - \sh(2 \om_1 - \la_1 - \la_2 - \h) \bigr)
              + \ch(\la_1 - \la_2) \sh(2 (\om_1 - \om_2))}
             {4 \sh(\la_1 - \la_2) \sh(\om_1 - \om_2 + \h)
	      \sh(\om_1 - \om_2 - \h)} \epp
\end{multline*}
Then
\begin{equation} \label{dppj1j2}
     D^{++}_{++} (\la_1, \la_2) = J_1 (\la_1, \la_2)
        + \lim_{\a \rightarrow 0} J_2 (\la_1, \la_2; \a) \epc
\end{equation}
where
\begin{align*}
     J_1 (\la_1, \la_2) & = 
        \biggl[ \prod_{j = 1}^2 \int_{\cal C}
	   \frac{\rd \om_j}{2 \p \i (1 + \fa(\om_j))}
	   \biggr]
        \frac{\det \bigl[ G(\om_j, \la_k; 0) \bigr]
	      \ch(\h) \sh(2 \om_2 - \la_1 - \la_2 - \h)}
             {2 \sh(\om_1 - \om_2 + \h) \sh(\om_1 - \om_2 - \h)
	     \sh(\la_1 - \la_2)} \epc \\
     J_2 (\la_1, \la_2; \a) & =
        \biggl[ \prod_{j = 1}^2 \int_{\cal C}
	   \frac{\rd \om_j \; \re^{- \a \h}}{2 \p \i (1 + \fa(\om_j))}
	   \biggr]
        \frac{\det \bigl[ G(\om_j, \la_k; \a) \bigr]
	      \cth(\la_1 - \la_2) \sh(2 (\om_1 - \om_2))}
             {4 \sh(\om_1 - \om_2 + \h) \sh(\om_1 - \om_2 - \h)} \epp
\end{align*}
Here $J_1 (\la_1, \la_2)$ is of a form which allows us to carry out
one integration by means of (\ref{G}) (for $\a = 0$). The result is
\begin{multline} \label{j1}
     J_1 (\la_1, \la_2) =
        \2 - \4 \bigl(\ph(\la_1; 0) + \ph(\la_2; 0)\bigr)
        - \4 \cth(\h) \ps(\la_1, \la_2; 0) \\
	+ \2 \cth(\la_1 - \la_2) \sum_{P \in \mathfrak{S}^2} \sign (P)
          \int_{\cal C}
	  \frac{\rd \om \; G(\om, \la_{P1}; 0) \cth(\om - \la_{P2} - \h)}
	       {2 \p \i (1 + \fa(\om))} \epp
\end{multline}
For the calculation of $J_2 (\la_1, \la_2; \a)$ we express the
hyperbolic functions in the integrand in terms of the kernel
(\ref{kernel}) occurring in the integral equation (\ref{G}) for $G$,
\begin{equation}
        \frac{\sh(2 (\om_1 - \om_2))}
             {\sh(\om_1 - \om_2 + \h) \sh(\om_1 - \om_2 - \h)} =
        \frac{K (\om_1 - \om_2; \a) - K (\om_2 - \om_1; \a)}
	     {2 \sh(\a \h)} \epp
\end{equation}
Then the integral over $\om_2$ can be performed by means of the
integral equation (\ref{G}) for finite $\a$, and we obtain
\begin{multline} \label{j2alpha}
     J_2 (\la_1, \la_2; \a) = - \4 \re^{- 2 \a \h} \cth(\la_1 - \la_2)
        \frac{\ps (\la_1, \la_2; \a) - \ps(\la_2, \la_1; \a)}
	     {2 \sh(\a \h)} \\
	- \2 \re^{- 2 \a \h} \cth(\la_1 - \la_2)
	  \sum_{P \in \mathfrak{S}^2} \sign (P)
          \int_{\cal C}
	  \frac{\rd \om \; G(\om, \la_{P1}; \a) \cth(\om - \la_{P2} - \h)}
	       {2 \p \i (1 + \fa(\om))} \epp
\end{multline}
From the definition (\ref{psi}) of $\ps$ and from the integral equation
(\ref{G}) for $G$ we infer the symmetry property $ \ps (\la_2, \la_1; \a)
= \ps(\la_1, \la_2; - \a)$ which can be used to carry out the limit
$\a \rightarrow 0$ for $J_2$. Using it and inserting the $\a \rightarrow
0$ limit of (\ref{j2alpha}) and (\ref{j1}) into (\ref{dppj1j2}) we
arrive at
\begin{multline} \label{dpp2}
     D^{++}_{++} (\la_1, \la_2) \\ = 
        \2 - \4 \bigl(\ph(\la_1; 0) + \ph(\la_2; 0)\bigr)
        - \frac{\cth(\h)}{4} \ps(\la_1, \la_2; 0)
        - \frac{\cth(\la_1 - \la_2)}{4 \h} \ps' (\la_1, \la_2; 0) \epc
\end{multline}
where the prime denotes the derivative with respect to $\a$. Inserting
now (\ref{dpmmp}), (\ref{dqpsi}) and (\ref{dpp2}) into (\ref{d2expand})
and taking into account the definitions (\ref{om}) and (\ref{defomprime})
of $\om$ and $\om'$ the reader will readily reproduce the density
matrix (\ref{D2res})-(\ref{D2odd}) for $n = 2$.

\Appendix{The high temperature expansions}
\label{app:hte}
\noindent
We comment on the application of high temperature expansions (HTE)
to the multiple integral formula, which provide important data for
the construction of the conjectures in this report. This may also be
a basis for the numerical evaluation of correlations as demonstrated in
\cite{TsSh05,Tsuboi07}.

As is usual, we assume an expansion of quantities in regular powers
of $\frac{1}{T}$. We then typically face the problem of solving a
linear integral equation for a unknown function $f(\la)$,
\begin{equation} \label{linearHT}
f(\la)=f_0(\la) +
 \nu \int_{\cal C} \frac{d\om}{2\pi i} K(\la-\om;\alpha) f(\om)
 := f_0(\la) + \nu K *f (\la) \epc
\end{equation}
where $\nu$ stands for some constant. 
The driving term  $f_0(\la)$ is a known function which has at most
simple poles at $\la=\mu_i$ and a pole of certain order at $\la=0$
inside ${\cal C}$.
Eq.\ (\ref{linearHT}) can be solved in an iterative manner,
\begin{equation}\label{iterative}
f(\la)=f_0(\la)+  \nu K *f_0 (\la)+
\nu^2 K* (K *f_0) (\la)+\cdots \epp
\end{equation}
The crucial observation is that $K *f_0 (\la)$ 
has poles at $\la=\pm \eta, \mu_i\pm \eta$
and  that these poles are outside of contour ${\cal C}$.
Thus, only the first two terms in (\ref{iterative}) do not vanish and 
$f(\la)=f_0(\la)+  \nu K *f_0 (\la)$ solves eq.\ (\ref{linearHT}).

This mechanism makes it possible to evaluate each order in the HTE in
an analytic and exact manner.
Of course, the evaluation of residues becomes more and more involved with
increasing order of $\frac{1}{T}$. 
Computer programs like Mathematica, however, can efficiently cope with
such a task and we obtain sufficiently many data for our purpose.

Here we present some examples which one can compute by hand.
We consider the nonlinear integral equation (\ref{nlie}) under the
assumption
$$
 \fa (\la) = 1+\frac{ \fa^{(1)} (\la)}{T}+\frac{ \fa^{(2)} (\la)}{T^2}+ \cdots \epp
$$
Then comparing  $O(\frac{1}{T})$ terms, one obtains the equation,
$$
\fa^{(1)}(\la) = a_0(\la) -\frac{1}{2}
 \int_{\cal C} \frac{d\om}{2\pi i} K(\la-\om;0) \fa^{(1)}(\om) \epc \qquad
a_0(\la)=- h - \frac{2J \sh^2 (\h)}{ \sh(\la) \sh(\la + \h)} \epp
$$
We apply the above strategy and find the first thermal correction to
$\fa(\la)$ as
$$
\fa^{(1)}(\la) = -h + 
\frac{2 J \sh^3(\h) \ch (\la)}{\sh(\la) \sh(\la-\h) \sh(\la+\h)}
$$
Similarly the first correction to $\bar{\fa}(\la)$ is found to be
$\bar{\fa}^{(1)}(\la) =-\fa^{(1)}(\la)$.

Equation (\ref{G}) can be solved similarly.
Let
$G(\la,\mu;\al) = G^{(0)}(\la,\mu;\al) + G^{(1)}(\la,\mu;\al)/T+\cdots$.
Then the following explicit forms are obtained.
\begin{align*}
 G^{(0)}(\la,\mu;\al)=&-\coth(\la-\mu)
                      + \re^{\al\eta}\frac{ \coth(\la-\mu-\h)}{2}
                    + \re^{-\al\eta}\frac{ \coth(\la-\mu+\h)}{2} \epc \\
 G^{(1)}(\la,\mu;\al)=&   - h \frac{K(\la-\mu;\al)}{4  }+
 \frac{ J \sh^3(\h) \ch (\mu)K(\la-\mu;\al) }
          {2 \sh(\mu) \sh(\mu-\h) \sh(\mu+\h)}  \\
 &+\frac{ J \sh \h K(\la;\al) G^{(0)}(0,\mu;\al) } {2} \epp
\end{align*}
%
%
%

All elements of the density matrix can now
be evaluated up to $O(T^{-1})$. 
A simple example is the emptiness formation probability for $n=2$,
\begin{align*}
D^{++}_{++}(\la_1,\la_2|T,h;0) &=\frac{1}{4}+
\frac{\fa^{(1)}(\la_1)+\fa^{(1)}(\la_2)}{8 T}
- \frac{J \sh \h G^{(0)}(0,\la_2;0)}{4 T}
 \frac{\sh \la_1 \sh(\la_2-\la_1+\h)}{\sh(\la_2-\la_1)\sh(\la_1-\h)} \\
&- \frac{J \sh \h G^{(0)}(0,\la_1;0)}{4 T}
 \frac{\sh \la_2 \sh(\la_1-\la_2+\h)}{\sh(\la_1-\la_2)\sh(\la_2-\h)} +O(T^{-2}) \epp
\end{align*}
The other basic functions are also readily evaluated.
\begin{align*}
 \varphi(\mu;\al)=&
 -\frac{h}{2T} +
 \frac{J \sh \h}{2T} \bigl(
  (1-\re^{-\al\eta}) \coth(\mu-\h)+ (1-\re^{\al\eta}) \coth(\mu+\h)
  \bigr) +O(T^{-2}) \epc \\
\psi(\mu_1,\mu_2;\al)=&
-\frac{1}{2} K(\mu_1-\mu_2;-\al)+
\frac{(\fa^{(1)}(\mu_2)- \fa^{(1)}(\mu_1)) G^{(0)}(\mu_2,\mu_1;\alpha)}
{2 T}\\
&-
\frac{J \sh \h}{T}  G^{(0)}(0,\mu_1;\al) G^{(0)}(\mu_2,0;\al) +O(T^{-2})
\epp
\end{align*}

One can then check the validity of our conjecture by comparing the
multiple integral formula for the density matrix and the exponential
formula after substitution of the basic functions by their HTE data up to 
$O(T^{-1})$. The higher order terms can, in principle, be checked
in the same manner.

Before closing the paper, we sketch briefly how we used
the HTE data to arrive at our conjecture.
Each density matrix element consists of two parts;
$D^{\rm even} $, the even part with respect to the magnetic field,
and $D^{\rm odd} $, the odd part.
The factorization for $n=2$ can be done fully in an analytic manner,
as demonstrated in appendix
\ref{app:doubleint}.
This result and the previous results of the XXX case motivate  us to
assume that the even part shares the same algebraic part 
with the ground state case.
Then it is not difficult to identify two basic functions,
$\phi(\la_1,\la_2), \tilde{\phi}(\la_1,\la_2)$.
We can actually represent them by the single function
$\psi(\la_1,\la_2;\al)$ such that  $\phi(\la_1,\la_2)= \frac{\sh \h}{2} \psi(\la_1,\la_2;0)$, $\tilde{\phi}(\la_1,\la_2)= -\frac{\sh \h }{2\h}
\frac{\partial}{\partial \al} \psi(\la_1,\la_2;\al)|_{\al=0}$.
This is one of the
advantages in  using the disorder parameter $\al$.

We then  consider the odd part of $n=3$.
The most interesting sector is 
$D^{\rm odd}_{3}\phantom{ }^{\e_1,\e_2,\e_3}_{\s_1,\s_2,\s_3}(\la_1,\la_2,\la_3)$
with $\sum_i \e_i = \sum_i \s_i=1$ to which  nine elements belong.
With use of the Yang-Baxter relation and the intrinsic symmetry
of the density matrix, one can represent all the element
by only one element.
We choose $D^{\rm odd}_{3}\phantom{ }^{++-}_{-++}$ for this, 
with permutations of the arguments $(\la_1,\la_2,\la_3)$.
The resulting 6 objects are found to satisfy linear algebraic relations,
and the consideration of the kernel space implies the representation
$$
D^{\rm odd}_{3}\phantom{ }^{++-}_{-++}(\x_1,\x_2,\x_3) = 
\frac{s_1(\x_1,\x_2,\x_3)}{\x_2} + \frac{s_2(\x_1,\x_2,\x_3)}{\x_1 \x_3} 
\epc
$$
where $\x_i=\re^{\la_i}$, and $s_1, s_2$ denote
certain symmetric functions of $\x_i$. 

We then assume that $s_1, s_2$ are given by sums of products of rational
functions of $x_i$ and the basic functions $\ph$, $\phi$ and
$\tilde{\phi}$, e.g.,
\begin{align*}
s_1(\x_1,\x_2,\x_3)=&
V_0(\x_1,\x_2|\x_3) \ph(\x_3;0)+
V_1(\x_1,\x_2|\x_3) \ph(\x_3;0) \phi(\x_1,\x_2)\\
& +
\tilde{V}_1(\x_1,\x_2|\x_3) \ph(\x_3;0) \tilde{\phi}(\x_1,\x_2) +
{\text{ cyclic permutations}}.
\end{align*}
$V_j(\x_1,\x_2|\x_3)$ is symmetric in $\x_1,\x_2$  ($j=0,1$) while 
$\tilde{V}_1(\x_1,\x_2|\x_3)$ is  anti-symmetric.

Furthermore, we restrict the possible forms of these coefficients
according to our previous experience such that

\begin{align*}
V_0(\x_1,\x_2|\x_3)&= \frac{p_0(\x_1,\x_2|\x_3)}{(\x_1^2-\x_3^2)(\x_2^2-\x_3^2)} \epc &
V_1(\x_1,\x_2|\x_3)&= \frac{p_1(\x_1,\x_2|\x_3)}{(\x_1^2-\x_3^2)(\x_2^2-\x_3^2)} \epc  \\
\tilde{V}_1(\x_1,\x_2|\x_3)&= 
\frac{\tilde{p}_1(\x_1,\x_2|\x_3)}{(\x_1^2-\x_3^2)(\x_2^2-\x_3^2)(\x_1^2-\x_2^2)} \epp
\end{align*}

Now $p_j$  and $\tilde{p}_1$ are polynomials in $\x_i$ and  $p_j$ ($\tilde{p}_1$ ) is 
symmetric (anti-symmetric) in  $\x_1,\x_2$.
We then assume polynomials of certain orders with desired
symmetry for them and fix the unknown coefficients so as to
match the HTE data.  All these parameters are 
fortunately fixed by the data up to $O(T^{-3})$.
At the final stage,  several hundreds of terms 
are canceled just by fixing one parameter, which looks
rather convincing.
We then check that the choice of parameters actually recovers the
$O(T^{-4})$ terms in the HTE.
After this procedure, we arrive at expressions for the
density matrix elements now written in terms of 
rational functions and basic functions. We then try to 
fit them into the exponential formula.
This requires the new operator $\mathbf{H}$ in the main body.
Once it is identified, it is easy to write down
the conjecture for $n=4$. Then again we test the validity against
the HTE data for the multiple integral formulae of the density matrix.

}


\begin{thebibliography}{10}

\bibitem{BGKS06}
H.~Boos, F.~G\"ohmann, A.~Kl\"umper and J.~Suzuki, \emph{Factorization of
  multiple integrals representing the density matrix of a finite segment of the
  {H}eisenberg spin chain}, J. Stat. Mech.  (2006) P04001.

\bibitem{BJMST04a}
H.~Boos, M.~Jimbo, T.~Miwa, F.~Smirnov and Y.~Takeyama, \emph{A recursion
  formula for the correlation functions of an inhomogeneous {XXX} model},
  Algebra and Analysis \textbf{17} (2005) 115.

\bibitem{BJMST05a}
---, \emph{Traces of the {S}klyanin algebra and correlation functions of the
  eight vertex model}, J. Phys. A \textbf{38} (2005) 7629.

\bibitem{BJMST06}
---, \emph{Algebraic representation of correlation functions in integrable spin
  chains}, Ann. Henri Poincar\'e \textbf{7} (2006) 1395.

\bibitem{BJMST05b}
---, \emph{Density matrix of a finite sub-chain of the {H}eisenberg
  anti-ferromagnet}, Lett. Math. Phys. \textbf{75} (2006) 201.

\bibitem{BJMST06b}
---, \emph{Hidden {G}rassmann structure in the {XXZ} model}, preprint,
  hep-th/0606280 (2006).

\bibitem{BJMST04b}
---, \emph{Reduced $q${KZ} equation and correlation functions of the {XXZ}
  model}, Comm. Math. Phys. \textbf{261} (2006) 245.

\bibitem{BJMST07app}
---, \emph{Fermionic basis for space of operators in the {XXZ} model},
  preprint, hep-th/0702086 (2007).

\bibitem{BoKo01}
H.~E. Boos and V.~E. Korepin, \emph{Quantum spin chains and {R}iemann zeta
  function with odd arguments}, J. Phys. A \textbf{34} (2001) 5311.

\bibitem{BoKo02}
---, \emph{Evaluation of integrals representing correlations in the {XXX}
  {H}eisenberg spin chain}, in M.~Kashiwara and T.~Miwa, eds., \emph{MathPhys
  Odyssey 2001 -- Integrable Models and Beyond -- In Honnor of Barry M. McCoy},
  65--108 (Birkh\"auser, Boston, 2002). Progress in Mathematical Physics, Vol.\
  23.

\bibitem{BKNS02}
H.~E. Boos, V.~E. Korepin, Y.~Nishiyama and M.~Shiroishi, \emph{Quantum
  correlations and number theory}, J. Phys. A \textbf{35} (2002) 4443.

\bibitem{BKS03}
H.~E. Boos, V.~E. Korepin and F.~A. Smirnov, \emph{Emptiness formation
  probability and quantum {K}nizhnik-{Z}amolodchikov equation}, Nucl. Phys. B
  \textbf{658} (2003) 417.

\bibitem{BKS04a}
---, \emph{New formulae for solutions of quantum {K}nizhnik-{Z}amolodchikov
  equation on level $-4$}, J. Phys. A \textbf{37} (2004) 323.

\bibitem{BKS04c}
---, \emph{New formulae for solutions to quantum {K}nizhnik-{Z}amolodchikov
  equations of level $-4$ and correlation functions}, Moscow Math. J.
  \textbf{4} (2004) 593.

\bibitem{BST05}
H.~E. Boos, M.~Shiroishi and M.~Takahashi, \emph{First principle approach to
  correlation functions of spin-1/2 {H}eisenberg chain: fourth-neighbor
  correlators}, Nucl. Phys. B \textbf{712} (2005) 573.

\bibitem{BHMMOZ06}
S.~Boukraa, S.~Hassani, J.-M. Maillard, B.~M. McCoy, W.~P. Orrick and
  N.~Zenine, \emph{Holonomy of the {I}sing model form factors}, J. Phys. A
  \textbf{40} (2007) 75.

\bibitem{DGHK07}
J.~Damerau, F.~G\"ohmann, N.~P. Hasenclever and A.~Kl\"umper, \emph{Density
  matrices for finite segments of {H}eisenberg chains of arbitrary length}, J.
  Phys. A \textbf{40} (2007) 4439.

\bibitem{FrRe92}
I.~B. Frenkel and {N.\ Yu.\ Reshetikhin}, \emph{Quantum affine algebras and
  holonomic difference equations}, Comm. Math. Phys. \textbf{1} (1992) 146.

\bibitem{GHS05}
F.~G\"ohmann, N.~P. Hasenclever and A.~Seel, \emph{The finite temperature
  density matrix and two-point correlations in the antiferromagnetic {XXZ}
  chain}, J. Stat. Mech.  (2005) P10015.

\bibitem{GKS04a}
F.~G\"ohmann, A.~Kl\"umper and A.~Seel, \emph{Integral representations for
  correlation functions of the {XXZ} chain at finite temperature}, J. Phys. A
  \textbf{37} (2004) 7625.

\bibitem{GKS05b}
---, \emph{Emptiness formation probability at finite temperature for the
  isotropic {H}eisenberg chain}, Physica B \textbf{359-361} (2005) 807.

\bibitem{GKS05}
---, \emph{Integral representation of the density matrix of the {XXZ} chain at
  finite temperature}, J. Phys. A \textbf{38} (2005) 1833.

\bibitem{GoSe05}
F.~G\"ohmann and A.~Seel, \emph{{XX} and {I}sing limits in integral formulae
  for finite temperature correlation functions of the {XXZ} chain}, Theor.
  Math. Phys. \textbf{146} (2006) 119.

\bibitem{JMMN92}
M.~Jimbo, K.~Miki, T.~Miwa and A.~Nakayashiki, \emph{Correlation functions of
  the {XXZ} model for {$\Delta < - 1$}}, Phys. Lett. A \textbf{168} (1992) 256.

\bibitem{JiMi95}
M.~Jimbo and T.~Miwa, \emph{Algebraic Analysis of Solvable Lattice Models}
  (American Mathematical Society, 1995).

\bibitem{JiMi96}
---, \emph{Quantum {KZ} equation with $|q| = 1$ and correlation functions of
  the {XXZ} model in the gapless regime}, J. Phys. A \textbf{29} (1996) 2923.

\bibitem{KSTS03}
G.~Kato, M.~Shiroishi, M.~Takahashi and K.~Sakai, \emph{Next-nearest-neighbour
  correlation functions of the spin-1/2 {XXZ} chain at the critical region}, J.
  Phys. A \textbf{36} (2003) L337.

\bibitem{KSTS04}
---, \emph{Third-neighbour and other four-point correlation functions of
  spin-1/2 {XXZ} chain}, J. Phys. A \textbf{37} (2004) 5097.

\bibitem{KMT99b}
N.~Kitanine, J.~M. Maillet and V.~Terras, \emph{Correlation functions of the
  {XXZ} {H}eisenberg spin-$\frac{1}{2}$ chain in a magnetic field}, Nucl. Phys.
  B \textbf{567} (2000) 554.

\bibitem{Kluemper92}
A.~Kl\"umper, \emph{Free energy and correlation length of quantum chains
  related to restricted solid-on-solid lattice models}, Ann.\ Physik \textbf{1}
  (1992) 540.

\bibitem{Kluemper93}
---, \emph{Thermodynamics of the anisotropic spin-1/2 {H}eisenberg chain and
  related quantum chains}, Z. Phys. B \textbf{91} (1993) 507.

\bibitem{SSNT03}
K.~Sakai, M.~Shiroishi, Y.~Nishiyama and M.~Takahashi, \emph{Third-neighbor
  correlators of a one-dimensional spin-1/2 {H}eisenberg antiferromagnet},
  Phys. Rev. E \textbf{67} (2003) 065101.

\bibitem{SaSh05}
J.~Sato and M.~Shiroishi, \emph{Fifth-neighbour spin-spin correlator for the
  anti-ferromagnetic {H}eisenberg chain}, J. Phys. A \textbf{38} (2005) L405.

\bibitem{SST05}
J.~Sato, M.~Shiroishi and M.~Takahashi, \emph{Correlation functions of the
  spin-1/2 anti-ferromagnetic {H}eisenberg chain: exact calculation via the
  generating function}, Nucl. Phys. B \textbf{729} (2005) 441.

\bibitem{Smirnov92b}
F.~A. Smirnov, \emph{Dynamical symmetries of massive integrable models. 1.
  {F}orm factor bootstrap equations as a special case of deformed
  {K}nishnik-{Z}amolodchikov equations}, Int. J. Mod. Phys. A \textbf{7} (1992)
  S813.

\bibitem{SAW90}
J.~Suzuki, Y.~Akutsu and M.~Wadati, \emph{A new approach to quantum spin chains
  at finite temperature}, J. Phys. Soc. Jpn. \textbf{59} (1990) 2667.

\bibitem{Suzuki85}
M.~Suzuki, \emph{Transfer-matrix method and {Monte Carlo} simulation in quantum
  spin systems}, Phys. Rev. B \textbf{31} (1985) 2957.

\bibitem{SuIn87}
M.~Suzuki and M.~Inoue, \emph{The {ST}-transformation approach to analytic
  solutions of quantum systems. {I}.\ {G}eneral formulations and basic limit
  theorems}, Prog. Theor. Phys. \textbf{78} (1987) 787.

\bibitem{Takahashi77}
M.~Takahashi, \emph{Half-filled {Hubbard} model at low temperature}, J. Phys. C
  \textbf{10} (1977) 1289.

\bibitem{TKS04}
M.~Takahashi, G.~Kato and M.~Shiroishi, \emph{Next nearest-neighbor correlation
  functions of the spin-1/2 {XXZ} chain at massive region}, J. Phys. Soc. Jpn.
  \textbf{73} (2004) 245.

\bibitem{Tsuboi07}
Z.~Tsuboi, \emph{A note on the high temperature expansion of the density matrix
  for the isotropic {H}eisenberg chain}, Physica A \textbf{377} (2007) 95.

\bibitem{TsSh05}
Z.~Tsuboi and M.~Shiroishi, \emph{High temperature expansion of the emptiness
  formation probability for the isotropic {H}eisenberg chain}, J. Phys. A
  \textbf{38} (2005) L363.

\end{thebibliography}

\end{document}